\documentclass[12pt]{article}

\usepackage[top=80pt,bottom=85pt,left=50pt,right=50pt]{geometry}
\usepackage{amssymb,amsmath,xypic}
\usepackage{slashed}
\usepackage{graphicx,color,subfigure}
\usepackage{float}
\usepackage{sectsty}
\usepackage{cite}
\usepackage[debug,pageanchor=false]{hyperref}
\definecolor{link}{rgb}{.8,.15,.1}
\hypersetup{colorlinks=true,linkcolor=link,citecolor=link,urlcolor=link,linktocpage}

\makeatletter
\@addtoreset{equation}{section}
\makeatother

\newcommand{\beq}{\begin{equation}}
\newcommand{\eeq}{\end{equation}}
\newcommand{\bea}{\begin{eqnarray}}
\newcommand{\eea}{\end{eqnarray}}
\newcommand{\nn}{\nonumber}

\begin{document}
%\begin{center}
%{\Large \bf{ Generalisation AdS$_3\times$S$^2$  }}
%\end{center}

\begin{titlepage}
	
	\begin{center}
		
		\vskip .5in %.3in 
		\noindent

		{\Large \bf{AdS$_3\times$ S$^2$ in IIB with  small ${\cal N}=(4,0)$ supersymmetry  } 
		 }

		\bigskip\medskip
 Niall T. Macpherson$^{a,}$\footnote{ntmacpher@gmail.com }, Anayeli Ramirez$^{b,c,}$\footnote{ramirezanayeli.uo@uniovi.es} \\

\bigskip\medskip
{\small 
	
	$a$: Departamento de F\'isica de Part\'iculas Universidade de Santiago de Compostela\\
	and\\
	Instituto Galego de F\'isica de Altas Enerx\'ias (IGFAE),\\
	R\'ua de Xoaqu\'in D\'iaz de R\'abago s/n,
	E-15782 Santiago de Compostela, Spain}\vskip 3mm
	
	$b$: Department of Physics, University of Oviedo,\\
	Avda. Federico Garcia Lorca s/n, 33007 Oviedo, Spain\vskip 3mm
	
	$c$: Kavli Institute for Theoretical Physics,\\
	 University of California, Santa Barbara, CA 93106

		\vskip 1.5cm 
		\vskip .9cm %.6cm
		{\bf Abstract }

		\vskip .1in
	\end{center}
	
	\noindent
We consider warped AdS$_3\times $S$^2\times$M$_5$ backgrounds in type II supergravity preserving small ${\cal N}=(4,0)$ supersymmetry. We show that imposing ${\cal N}=(4,0)$ supersymmetry imposes between 0 and 3 a priori isometries in the internal M$_5$. In this work we focus on classes of solution where M$_5$ exhibits no a priori  isometry which imposes additional constraints. Solving these in IIA forces M$_5$ to support an SU(2)-structure, a class already studied in \cite{Lozano:2019emq}, while in IIB one arrives at two broad new classes with identity-structure that we reduce to local expressions for the physical fields and PDEs.
	
	\noindent
	
	\vfill
	\eject
	
\end{titlepage}

\tableofcontents
\section{Introduction and summary}
Two dimensional superconformal algebras  come in a  wide variety of different types \cite{Fradkin:1992bz} which should be contrasted with their higher dimensional counterparts. The classification and construction of supersymmetric AdS$_3$ string vacua realising these algebras\footnote{The algebras that can be embedded into $d=10/11$ supergravity are classified in \cite{Beck:2017wpm}} is a rich topic that is still mostly unknown. This is unfortunate because such solutions have rather broad applications with relevance to the  AdS$_3$/CFT$_2$ correspondence, duals to surface defects in higher dimensional SCFTs and the near horizons of  black strings.  A particular case of some importance are small ${\cal N}=(4,0)$ AdS$_3$ vacua in 10 dimensions. The construction and classification of these is the focus of this work. See \cite{Maldacena:1997de,Kim:2007hv,OColgain:2010wlk,Lozano:2015bra,Couzens:2017way,Dibitetto:2017tve,Dibitetto:2018iar,Lozano:2019emq,Lozano:2019jza,Lozano:2019zvg,Lozano:2019ywa,Faedo:2020lyw,Lozano:2020bxo,Dibitetto:2020bsh,Faedo:2020nol,Zacarias:2021pfz,Couzens:2021veb,Couzens:2020aat}  for related small ${\cal N}=(4,0)$  work\footnote{AdS$_2$ with small  ${\cal N}=(4,0)$ was also considered in 
\cite{Dibitetto:2018gtk,Lozano:2020txg,Lozano:2020sae,Lozano:2021rmk,Lozano:2021fkk,Ramirez:2021tkd}
} and 
\cite{Martelli:2003ki,Kim:2005ez,DHoker:2008lup,DHoker:2008rje,Donos:2008hd,Kelekci:2016uqv,Eberhardt:2017uup,Couzens:2017nnr,Dibitetto:2018ftj,Macpherson:2018mif,Legramandi:2019xqd,Couzens:2019iog,Passias:2020ubv,Legramandi:2020txf,Couzens:2021tnv,Macpherson:2021lbr,Edery:2020kof}
 for some works realising other algebras. \\
~\\
The small  ${\cal N}=(4,0)$ algebra is $\mathfrak{su}(1,1|2)/\mathfrak{u}(1)$ which has an  SU(2)  R-symmetry  and comes equipped with a multiplet of supercurrents in the $\textbf{2}\oplus \bar{\textbf{2}}$ representation of this group. To construct an AdS$_3$ solution realising this algebra it is necessary that its internal space M$_7$ realises the R-symmetry.  Specifically the bosonic supergravity fields should be SU(2) singlets while the internal spinors should transform in the $\textbf{2}\oplus \bar{\textbf{2}}$. This leads quite naturally to M$_7$ being a foliation of a 2-sphere over some M$_5$, which can be fibered over the S$^2$ provided SU(2) is preserved. In this work we shall assume M$_7=$ S$^2\times$M$_5$ is warped product which enables us to use a set of general SU(2) spinors already constructed on this space in \cite{Lozano:2019emq}. Under mild assumptions\footnote{Consistency with a non trivial Romans mass and the presence of simple D brane and O plane sources} IIA solution  on this space with M$_5$ supporting an SU(2)-structure were completely classified in \cite{Lozano:2019emq}, leading to an interesting proposal for a particular AdS$_3$/CFT$_2$ correspondence in \cite{Lozano:2019jza,Lozano:2019zvg,Lozano:2019ywa}. A main goal of this work is to move beyond SU(2)-structure and consider more generic classes of solutions where M$_5$ supports an identity-structure, with a view towards similar AdS$_3$/CFT$_2$ applications.\\
~~\\
Generalising from SU(2) to Identity-structure significantly complicates matters, the reason is many of the at least $\frac{1}{4}$ BPS superconformal algebras contain an R-symmetry for which SU(2) is a subgroup - the most obvious being the large ${\cal N}=(4,0)$ algebra which contains two copies of the  small algebra. It just so happens that these other algebras are inconsistent with the assumption of SU(2)-structure - for Identity-structure this is no longer the case. Rather than attempting to brute force ones way through the classification of all warped AdS$_3\times$ S$^2$ solutions, it would be beneficial to have some way to identify exactly what algebra a class of solution is realising before descending down a rabbit hole of computation. Another main motivation of this work is to provide precisely such a tool.\\
~~\\
The lay out of the paper is as follows:\\
~~\\
In section \ref{sec:relinsingsu2} we spell out how we realise small ${\cal N}=(4,0)$ supersymmetry for warped AdS$_3$ solutions of type II supergravity. We begin in section \ref{sec: ads3susy} by reviewing the necessary geometric conditions for ${\cal N}=(1,0)$ supersymmetric AdS$_3$ \cite{Dibitetto:2018ftj}. In section \ref{sec:neq4} we  explain how this may be used as a stepping stone to construct solutions with at least small ${\cal N}=(4,0)$ supersymmetry. We also give details of the ansatz we are taking, namely that the internal space decomposes as a foliation of the round  S$^2$ over M$_5$, and construct general spinors on this space spinors consistent with an SU(2) R-symmetry -  generically these give rise to an identity-structure on M$_5$. This section is supplemented by appendices \ref{sec:gammamatrices} and \ref{sec:derivationof5dconstraints} where we derive totally general geometric conditions on M$_5$ that imply ${\cal N}=(4,0)$ supersymmetry. In general these conditions are rather unwieldy, so in section  \ref{sec:exactlysmall} we introduce a new method to aid in the construction of AdS$_3$ solutions with extended supersymmetry: We introduce a matrix bilinear of Killing vectors which the spinors of an AdS$_3$ solution with at least ${\cal N}=(2,0)$ supersymmetry are necessarily charged under. This allows us to identify several things about a class of solutions a priori, first it makes clear under what conditions S$^2$ will experience an enhancement to S$^3$, second it tells us how many a priori isometries $(0,1,2,3)$ supersymmetry demands M$_5$ must contain, third it establishes exactly which algebra is being realised. We decide to focus specifically on the classes that realise exactly small ${\cal N}=(4,0)$ rather than large ${\cal N}=(4,0)$ or some other more supersymmetric algebra. We also focus on the cases where S$^2$ is not enhanced to S$^3$ because all such solutions can be generated with string dualities from solutions with round 2-spheres. We prove this in section \ref{sec:generalityofansatz} where we also comment on the generality, modulo duality, of assuming that the  S$^2$ realising the required SU(2) R-symmetry does not appear with additional U(1) isometries fibred over it as it could generically.\\
~\\
Finding all classes of solution on AdS$_3\times $S$^2\times$M$_5$ preserving small ${\cal N}=(4,0)$ supersymmetry is a significant undertaking. We begin this process in section  \ref{eq:noisometry}, by classifying solutions for which supersymmetry imposes no a priori isometries in M$_5$, as modulo duality these likely represent the most general classes. We leave the classification of solutions with 1--3 isometries in M$_5$ for future work. It turns out that imposing no  a priori isometries in IIA restricts the ansatz to SU(2)-structure, already considered in \cite{Lozano:2019emq},  as such the focus of the rest of this work will be on type IIB where M$_5$ necessarily supports an identity-structure. The  conditions for supersymmetry in appendix \ref{sec:derivationof5dconstraints} truncate considerably, and we are able to establish that there are 2 classes of solution: For class I (section \ref{sec:classI}) the Bianchi identity of the RR 1-form is implied by supersymmetry while for class II (section  \ref{sec:classII}) it is not, making only the latter compatible with co-dimension 2 sources (D7 branes etc).
 
We reduce the conditions for the existence of solutions in these classes to locally expressions for the supergravity fields and a set of PDEs which imply supersymmetry and the type IIB equations of motion. As this is an involved process, we begin by classifying a sub-class of class I in section \ref{sec:classI0}, which is simple enough for us to explicitly explain the methods we apply more broadly. Once the Bianchi identities are also considered this sub-class branches into 2 cases i) D5 branes ending on NS5 branes both wrapping AdS$_3\times$ S$^2$, ii) The T-dual of a IIA solution with a round 3-sphere.  We then study class I in full generality in section \ref{sec:classI}, where the Bianchi identities no longer impose an obvious branching of solutions generically. Clearly the general class is more complicated, however the governing PDEs are still reminiscent of intersecting brane scenarios. We take the T-dual of  a sub-case that yields a now squashed and fibred  3-sphere IIA class in section \ref{eq:S3class} which significantly generalises a class of solutions found in \cite{Faedo:2020nol}. 

In section \ref{sec:classII} we derive the second class of solutions with no a priori isometry in M$_5$. This class turns out to be significantly more involved, leading to a rather intimidating set of governing PDEs. Experience suggests to us that this indicates the class contains many physically distinct cases, and also that there are likely better local coordinates to express the system in terms of, at least once restrictions are made. We consider one such restriction in section \ref{eq:classIIrestriction} - that the metric is diagonal. In terms of a new set of coordinates we find two cases i) A deformed D5-NS5 brane intersection T-dual to a squashed and fibered 3-sphere. ii) A case with  no necessary isometry governed by a system of PDEs generalising  the D8-NS5-D6 Mink$_6$ system of \cite{Imamura:2001cr}, albeit this case is in IIB and with D7-NS5-D5 branes extended in AdS$_3\times$ S$^2$.  

Although we do not consider this here, the classification of solutions with no a priori isometry in M$_5$ should also be supplemented by the classification of solution with at least 1 isometry: Ultimately although supersymmetry does not, the Bianchi identities of the fluxes impose an isometry in M$_5$ in several of the cases we consider. It is possible that such cases are restrictions of more general classes of solution where supersymmetry does indeed impose an isometry. Also it is entirely possible that one needs to consider such classes to capture the T-dual of every round 3-sphere class. We shall return to these issues in \cite{NA}, where we shall also populate the classes of solution we find here.

\section{Realising small ${\cal N}=(4,0)$}\label{sec:relinsingsu2}
In this section we illiterate how classes of AdS$_3$ solutions in type II supergravity realising ${\cal N}=(4,0)$ can be constructed. We begin by reviewing some features of supersymmetric AdS$_3$ in general in section \ref{sec: ads3susy}.  In section \ref{sec:neq4} we find the general form of ${\cal N}=(4,0)$ spinors on warped AdS$_3\times$ S$^2\times$M$_5$ that transform in $2\oplus \overline{2}$ of SU(2) and are also consistent with physical fields that are SU(2) singlets. In section \ref{sec:exactlysmall} we introduce a method to analysis the isometry structure the spinors an AdS$_3$ solution imply, allowing us to focus on small ${\cal N}=(4,0)$ classes specifically. Finally in section \ref{sec:generalityofansatz} we explore the generality  of the AdS$_3\times$ S$^2\times$M$_5$ ansatz we make throughout this section. We comment on what happens when we allow M$_5$ to be fibred over S$^2$, and identify exactly what is not contained in our ansatz modulo duality. 

\subsection{Supersymmetric AdS$_3$ in type II supergravity}\label{sec: ads3susy}
We are interested in supersymmetric AdS$_3$ solutions of type II supergravity. As such we restrict our attention to solutions for which the bosonic fields decompose as
\beq\label{eq:AdS3bosonicfields}
ds^2= e^{2A} ds^2(\text{AdS}_3)+ ds^2(\text{M}_7),~~~~ H_{10}=c \text{vol}(\text{AdS}_3)+ H,~~~~F_{10}= f_{\pm}+ e^{3A}(\text{AdS}_3)\wedge \star_7\lambda (f_{\pm}),
\eeq
where in IIA $f_+=f_0+f_2+f_4+f_6$ or IIB $f_-=f_1+f_3+f_5+f_7$ is the magnetic part of the RR poly form $F_{10}$, $H_{10}$ is the NS 3-form and  $\lambda X_n= (-1)^{[\frac{n}{2}]}X_n$ for any n-form. The fields $(e^{A},f_{\pm},H_3)$ and the dilaton $\Phi$ have support on M$_7$ only and $c$ is a constant. The RR fluxes should obey $dF_{10}=H\wedge F_{10}$ away from sources, necessitating
\beq\label{eq:7dbianchis}
d_Hf_{\pm}=0,~~~~~ d_H(e^{3A}\star_7\lambda(f_{\pm}))= c f_{\pm},
\eeq
in regular parts of a solution, where we define the twisted derivative $d_H=d-H\wedge $. An immediate consequence is that in IIA $c f_0=0$ in general, so either the NS 3-form is purely magnetic, or there is no Romans mass. In IIB one can always exploit  SL(2,$\mathbb{R}$) duality to move to a duality frame with $c=0$. In either IIA or IIB if we assume only space-time filling sources, the magnetic flux Bianchi identity gets modified in their presence but the electric one does not: Taking $d_H$ of the later then implies that for $c\neq 0$, a RR source is only possible when an NS sources is also present at its loci - ie there can be no simple D brane or O plane sources when $c \neq 0$, only more exotic objects composite objects. For these reasons we shall fix
\beq
c=0,
\eeq
where more general solutions can be generated via duality, or, when they are in IIA, would be better studied from a d=11 perspective.\\
~~\\
When an AdS$_3$ solutions preserve at least ${\cal N}=(1,0)$ supersymmetry  it may be defined in terms of two real bi-spinors $\Psi_{\pm}$ \cite{Dibitetto:2018ftj}, themselves defined in terms of two $d=7$ Majorana spinors $\chi_{1,2}$ as
\beq
\Psi_++ i\Psi_- =  \chi_1\otimes \chi_2^{\dag},
\eeq
the RHS of this expression is defined in \eqref{bispinors}.  These bi-spinors are related to the supergravity fields by the geometric conditions\footnote{These conditions hold for $c=0$, the general conditions were only recently derived in \cite{Macpherson:2021lbr} }
\begin{subequations}
	\begin{align}
		&d_{H}(e^{A-\Phi}\Psi_{\mp})=0,\label{eq:bps1}\\[2mm]
		&d_{H}(e^{2A-\Phi}\Psi_{\pm})\mp 2m e^{A-\Phi}\Psi_{\mp}=\frac{e^{3A}}{8}\star_7\lambda(f_{\pm}),\label{eq:bps2}\\[2mm]
		&(\Psi_{-},f_{\pm})_7= \mp\frac{m}{2} e^{-\Phi}\text{vol}(\text{M}_7)\label{eq:bps3},
	\end{align}
\end{subequations}
where $\pm$ should be taken in IIA/IIB, we share the conventions of \cite{Passias:2020ubv} and $|\chi_1|^2=|\chi_2|^2= e^{A}$. These condition are necessary and sufficient for supersymmetry, but not to have a solution of type II supergravity in general, for that one needs to also impose the RR and NS flux Bianchi identities and that, if sources are present, they have a supersymmetric embedding - the remaining EOM are then implied. 
%%%%%%%%%%%%%%%%%%%%%%%%%%%%%%%%%%%%%%%%%%%%%%%%%%%%%%%%%%%%%%%%%%%%%%%%%%%%%%%
\subsection{An ansatz for at least small ${\cal N}=(4,0)$ supersymmetry}\label{sec:neq4}
To have ${\cal N}=(4,0)$ supersymmetry we must have 4 independent sets of Majorana spinors on the internal space $(\chi_1^I,\chi_2^I)$ for $I=1,...,4$, that each solve \eqref{eq:bps1}-\eqref{eq:bps3}  for the same bosonic fields, ie the same metric, dilaton and fluxes.  To realise small ${\cal N}=(4,0)$ specifically it is necessary that  $\chi_{1,2}^I$ transform in the $\textbf{2}\oplus\overline{\textbf{2}}$ representation of SU(2), the R-symmetry of small ${\cal N}=(4,0)$ -  the bosonic fields should be singlets under its action. This means that for $K_i$, $i=1,2,3$, the SU(2)
 Killing vectors we must have
\beq\label{eq:SU2lie}
{\cal L}_{K_i}\chi_{1,2}^I=\frac{i}{2}(\Sigma_i)^{IJ}\chi_{1,2}^J,~~~~ {\cal L}_{K_i}(A,\Phi,g(\text{M}_7),f,H_3)=0,
\eeq
where $\frac{i}{2}\Sigma_i$ span the $\textbf{2}\oplus\overline{\textbf{2}}$ of $\mathfrak{su}(2)$. This provides a map between each of the 4 ${\cal N}=1$ sub-sectors of $(\chi_1^I,\chi_2^I)$, and one can show that if a single one of these solves a set of sufficient conditions for  ${\cal N}=1$ supersymmetry the other 3 also necessarily solve these conditions \cite{Legramandi:2020txf}.  The non trivial part is constructing a set of spinors such that \eqref{eq:SU2lie} holds.\\
~~\\
Given that we need an SU(2) R-symmetry it should not be hard to convince oneself that M$_7$ needs to decompose in terms of a 2-sphere and some 5 manifold M$_5$. The 2-sphere could be the round one, or M$_5$ could be fibered over it such that SU(2) is preserved. We will make the ansatz that this 2-sphere is the round one, and discuss the generality of this assumption in section \ref{sec:generalityofansatz}. We shall thus refine \eqref{eq:AdS3bosonicfields} in terms of a unit radius 2-sphere as
\beq\label{eq:SU2ansatz}
ds^2(\text{M}_7)= e^{2C}ds^2(\text{S}^2)+ds^2(\text{M}_5),~~~H= e^{2C}H_1\wedge \text{vol}(\text{S}^2)+H_3,~~~~f_+= g_{1\pm}+e^{2C}g_{2\pm}\wedge \text{vol}(\text{S}^2),
\eeq
where $(e^{A},e^{C},\Phi,g_1,g_2,H_1,H_3)$ have support on M$_5$ alone which does not depend on the S$^2$ coordinates. A general set of Majorana SU(2) spinors transforming in the $\textbf{2}\oplus\overline{\textbf{2}}$ were already derived on this geometry in  \cite{Lozano:2019emq}, they are
\begin{align}
\chi^I_1&=\frac{e^{\frac{A}{2}}}{\sqrt{2}}({\cal M}^I)_{\alpha\beta} \bigg(\xi^{\alpha}\otimes\eta^{\beta}_{11}+i \sigma_3\xi^{\alpha}\otimes\eta^{\beta}_{12}\bigg),~~~\chi^I_2= \frac{e^{\frac{A}{2}}}{\sqrt{2}}({\cal M}^I)_{\alpha\beta}\bigg(\xi^{\alpha}\otimes\eta^{\beta}_{21}+i\sigma_3\xi^{\alpha}\otimes\eta^{\beta}_{22}\bigg),\label{eq:neq4spinor}
\end{align}
where ${\cal M}^I=(\sigma_2\sigma_1,\sigma_2\sigma_2,\sigma_2\sigma_3,-i \sigma_2)^I$. Here $\xi^{\alpha},\sigma_3\xi^{\alpha}$ are independent SU(2) doublets of Killing spinors on S$^2$ (see appendix \ref{sec:appS2}) and $\eta^{\alpha}_{ij}$ are two component vectors with spinorial entries: Specifically there are 4 independent spinors on M$_5$ namely $\eta_{11},\eta_{12},\eta_{21},\eta_{22}$ and $\eta^{\alpha}_{ij}$ depend on these and their Majorana conjugates $\eta^c_{ij}$ (see appendix \ref{sec:appM2}). The specific representation appearing in \eqref{eq:SU2lie} for these spinors is 
\beq
\Sigma_i=(\sigma_2\otimes \sigma_1,~-\sigma_2\otimes \sigma_3,~\mathbb{I}\otimes\sigma)_i,
\eeq
which one can confirm is equivalent to the $\textbf{2}\oplus \overline{\textbf{2}}$.
In what follows we shall take our ${\cal N}=1$ sub-sector to be
\beq
\chi_1=\chi_1^2,~~~~\chi_2=\chi_2^2\label{eq:neq1subsector}
\eeq
Inserting \eqref{eq:neq1subsector} into the supersymmetry conditions \eqref{eq:bps1}-\eqref{eq:bps3} leads to a set of 5d bi-linear constraints that we derive in appendix \ref{sec:derivationof5dconstraints} resulting in a sufficient but highly degenerate system of constraints \eqref{eq:IIfirstbispinorconds1}-\eqref{thirdgen} for IIA and \eqref{5deqIIB1}- \eqref{thirdgenIIB} in IIB.  These conditions are rather crude, and the main purpose of this and the next section is to refine them. First using \eqref{eq:neq1subsector} to compute $|\chi_{1,2}|^2$ it becomes apparent they generically depend on the SU(2) embedding coordinates $y_i$, which are SU(2) triplets. Fixing $|\chi_{1,2}|^2=e^{A}$ as supersymmetry demands requires that we impose
\begin{align}
\eta^{c\dag}_{12}\eta_{11}=\text{Im}(\eta^{\dag}_{12}\eta_{11})=\eta^{c\dag}_{22}\eta_{21}=\text{Im}(\eta^{\dag}_{22}\eta_{21})=0,\nn\\[2mm]
|\eta_{11}|^2+|\eta_{12}|^2=|\eta_{21}|^2+|\eta_{22}|^2=1\label{eq:eAsinglet}.
\end{align}
In order to solve these it is helpful to decompose the spinors in a common basis in terms of a single unit norm spinor $\eta$. Such a 5d spinor defines an SU(2)-structure in 5d as
\beq
\eta\otimes \eta^{\dag}= \frac{1}{4}(1+V)\wedge e^{-i J},~~~~\eta\otimes \eta^{c\dag}= \frac{1}{4}(1+V)\wedge\Omega,
\eeq
where $J$ is a (1,1)-form and $\Omega$ as (3,0)-form, they are defined on the sub-manifold M$_4\subset$ M$_5$ orthogonal to the real 1-form $V$, and obey
\beq
J^3=\frac{3i}{4}\Omega\wedge \overline{\Omega},~~~~~J\wedge \Omega=0.
\eeq
This leads to a natural decomposition of the internal 5-manifold as
\beq
ds^2(\text{M}_5)= V^2+  ds^2(\text{M}_4).
\eeq
We can decompose a generic spinor $\tilde{\eta}$ in terms of $\eta$, a holomorphic 1-form on M$_4$ $Z$ and some complex functions $p_1,p_2,p_3$ as
\beq
\tilde{\eta}= p_1 \eta+ p_2\eta^c+ \frac{|p_3|}{2} \overline{Z}\eta,~~~ Z\eta=0.
\eeq
Using these facts, and after a lengthy calculation one can show that a set of general 5d spinors solving \eqref{eq:eAsinglet} are given by
\begin{align}
\eta_{11}&=\sin\left(\frac{\alpha_1+\alpha_2}{2}\right)\eta,~~~\eta_{12}=\cos\left(\frac{\alpha_1+\alpha_2}{2}\right)(\cos\beta_1+ \sin\beta_1 \frac{1}{2}\overline{Z}_1)\eta\nn\\[2mm]
\eta_{21}&=\sin\left(\frac{\alpha_1-\alpha_2}{2}\right)\eta_1,~~~\eta_{22}=\cos\left(\frac{\alpha_1-\alpha_2}{2}\right)(\cos\beta_2 \eta_1+ \sin\beta_2 ( d_1\eta_2+d_2 \eta^c_2))\label{eq:genspinors}
\end{align}
where 
\beq
\eta_1= \sqrt{1-c^2}(a \eta+ b \eta^c)+ \frac{c}{2}\overline{Z}_2\eta,~~~\eta_2=c\big(a \eta+ b \eta^c\big)- \sqrt{1-c^2}\frac{1}{2}\overline{Z}_2\eta.
\eeq
Here we have to introduce two generic holomorphic 1-forms $Z_1,Z_2$, several real functions of M$_5$ $(\alpha_{1,2},\beta_{1,2}, c)$ for $|c|\leq 1$ and some complex ones constrained as 
\beq
|a|^2+|b|^2= |d_1|^2+|d_2|^2=1.
\eeq
We can assume that $\eta_{11}$ never vanishes without loss of generality, but the other spinors contain some redundancy when certain parts of the other spinors are turned off. For instance when $c=1$ we can fix $(d_1=|d_1|,d_2=0)$ without loss of generality, while when $\eta_{12}=0$ we may also fix $\beta_2=0$ without further cost. The presence of $(Z_1,Z_2)$ in \eqref{eq:genspinors} indicates that M$_4$ generically supports an identity-structure, however when $(c=0=\beta_1=\beta_2=0)$ the 1-forms  drop out and this becomes an SU(2)-structure- in IIA this case was already completely classified in \cite{Lozano:2019emq}. Generically $(Z_1,Z_2)$ are neither parallel nor orthogonal, rather in general they may be used to define 2 complex functions $(z,\tilde{z})$ as
\beq
z= \frac{1}{2}\iota_{\overline{Z}_1}Z_2,~~~~\tilde{z}= \frac{1}{4}\iota_{\overline{Z}_1}\iota_{\overline{Z}_2}\Omega,~~~~|z|^2+|\tilde{z}|^2=1.
\eeq
which vanish when  the 1-forms are respectively orthogonal and parallel.  Only when $z=0$ do  $(Z_1,Z_2)$ define a vielbein on M$_4$, otherwise one can assume that $(Z_1,\frac{1}{2}\iota_{Z_1}\Omega)$ do with out loss of generality with $Z_2$ defined along each of these.  At this point one can proceed to try to solve $d=5$ supersymmetry conditions derived in the appendix \ref{sec:derivationof5dconstraints}. A first import thing to note is that these contain several algebraic constraints: In IIA  \eqref{eq:IIfirstbispinorconds1}-\eqref{thirdgen} imply the following conditions
\beq\label{eq:zeroformsIIA}
\begin{split}
	&\eta_{21}^{c\dag}\eta_{11}=\eta_{22}^{c\dag}\eta_{12},\qquad \text{Im}(\eta_{21}^{\dag}\eta_{11})=\text{Im}(\eta_{22}^{\dag}\eta_{12}),\\
	&(1+2me^{C-A})\eta_{22}^{c\dag}\eta_{11}=(1-2me^{C-A})\eta_{21}^{c\dag}\eta_{12},\\
	&(1+2me^{C-A})\text{Im}(\eta_{22}^{\dag}\eta_{11})=(1-2me^{C-A})\text{Im}(\eta_{21}^{\dag}\eta_{12}).
\end{split}
\eeq
In IIB on the other hand \eqref{5deqIIB1}- \eqref{thirdgenIIB} imply
\beq\label{eq:zeroformsIIB}
\begin{split}
&\eta_{22}^{c\dag}\eta_{11}=\eta_{21}^{c\dag}\eta_{12},\qquad \text{Im}(\eta_{22}^{\dag}\eta_{11})=\text{Im}(\eta_{21}^{\dag}\eta_{12}),\\
&(1+me^{C-A})\eta_{21}^{\dag}\eta_{11}=(1-me^{C-A})\eta_{22}^{\dag}\eta_{12},\\
&(1+me^{C-A})\text{Im}(\eta_{21}^{c\dag}\eta_{11})=(1-me^{C-A})\text{Im}(\eta_{22}^{c\dag}\eta_{12}).
\end{split}
\eeq
Unfortunately plugging our spinor ansatz into the IIA or IIB algebraic conditions leads to a  lot of branching possibilities, and the constraints are rather intractable in general - though progress can be made making assumptions.  Rather than attempting a brute force approach it would be beneficial to have some additional guiding principle. Here it is opportune to make one point clear: There are rather a lot of superconformal algebras consistent with AdS$_3$, and many of those preserving at least 8 chiral super charges can admit solutions consistent with the ansatz taken so far. It makes sense to attempt to zoom in on those preserving the small algebra specifically, that is what we seek after all. Additionally what remains would be better tackled with a more specialised ansatz. For example, large ${\cal N}=4$ is consistent with the ansatz taken thus far, but for that we know M$_5$ must contain a 2 or 3-sphere and it simplify matters to assume its presence from the start.\\
~~\\
In the next section we shall narrow our focus to solutions that specifically realise the small ${\cal N}=(4,0)$ algebra rather than something larger.   
\subsection{Isolating the small algebra and a priori isometries in M$_5$} \label{sec:exactlysmall}
In this section we introduce a method to restrict the internal spinors of an AdS$_3$ solution with extended supersymmetry to those that realise a particular superconformal algebra, in this case  small ${\cal N}=(4,0)$, though let us stress that  this technique could be applied to any algebra with at least ${\cal N}=(2,0)$ extended supersymmetry.\\
~~\\
In \cite{Tomasiello:2011eb} generic supersymmetric solutions of type II supergravity are classified, one of their findings is that supersymmetry implies that the following $d=10$ bi-linear
\beq\label{eq:tenKill}
K^{(10)}= \frac{1}{64}(\overline{\epsilon}_1\Gamma^M\epsilon_1+\overline{\epsilon}_1\Gamma^M\epsilon_1)\partial_{M},
\eeq
 is a Killing vector with respect to all the bosonic fields. Additionally the  $d=10$ Majorana Weyl spinors $\epsilon_{1,2}$ are singlets with respect to it. Taking an ${\cal N}=(2,0)$ AdS$_3$ spinor ansatz for $\epsilon_{1,2}$ involving two sets of $d=7$ spinors $(\chi^1_{1,2},\chi^2_{1,2})$ it then necessarily follows that
\beq
K^{({\cal N}=2)}=-i(\chi^1_1(\gamma^{(7)})^a\chi^2_1 \mp  \chi^1_2(\gamma^{(7)})^a\chi^2_2)\partial_{a},~~~a=1,...,7\label{eq:neq2killing}
\eeq
 defines a Killing vector on the internal space under which $(\chi^1_{1,2},\chi^2_{1,2})$ are charged and the bosonic fields are singlets. The computation is similar to that appearing in \cite{Passias:2017yke} for ${\cal N}=2$ AdS$_4$, and  appears for the AdS$_3$ case in \cite{Couzens:2022agr}. For a set of  ${\cal N}=(n,0)$ spinors one can define $\frac{n}{2}(n-1)$ such ${\cal N}=(2,0)$ sub-sectors, so it follows that
\beq
K^{IJ}= -i (\chi^{I\dag}_1(\gamma^{(7)})^a\chi^J_1 \mp  \chi^{I\dag}_2(\gamma^{(7)})^a\chi^J_2)\partial_{a},\label{eq:extendedKillingvectors}
\eeq
defines an antisymmetric matrix of Killing vectors under which the bosonic fields are singlets, and $\chi_{1,2}^I$ are charged. The entries of $K^{IJ}$ (modulo antisymmetry) are not however necessarily independent\footnote{If it is not clear this is not a weakness: Small ${\cal N}=(4,0)$ comes with 3 independent Killing vectors, a generic antisymmetric $4\times 4$ matrix 6 independent components, so we need half of them to be dependent in this case.}. It is thus natural to identify $K^{IJ}$ with the Killing vectors associated to the R-symmetry, this is almost correct, but generically $K^{IJ}$ could be a linear combination of these and a number of flavour isometries.\\
~\\
We shall now turn our attention to our ${\cal N}=(4,0)$ spinors \eqref{eq:neq4spinor}. For these \eqref{eq:extendedKillingvectors} must decompose in terms of vector bi-linears on S$^2$ and M$_5$, one can show that the former are
\beq
\xi^{\alpha\dag}\xi^{\beta}= \delta^{\alpha\beta},~~~\xi^{\alpha\dag}\sigma_3\xi^{\beta}= - y_i(\sigma_i)^{\alpha\beta},~~~~\xi^{\alpha\dag}\gamma_2^{\mu}\xi^{\beta}\partial_{\mu}= (k_i)^{\mu}\partial_{\mu}(\sigma_i)^{\alpha\beta},~~~~\xi^{\alpha\dag}\gamma_2^{\mu}\sigma_3\xi^{\beta}\partial_{\mu}=-i (dy_i)^{\mu}\partial_{\mu}(\sigma_i)^{\alpha\beta},\nn
\eeq
where $\gamma_2^{\mu}$ are curved 2d gamma matrices, $K_i=\epsilon_{ijk}dy_j y_k$ are the 1-forms dual to the SU(2) Killing vectors on S$^2$ and $y_i$ are unit norm embedding coordinates. The relevant 5d bi-linears can be expressed in terms of the following 0 and 1-form bi-linears 
\begin{align}
{\cal F}_1=&2e^{A-C}(\eta_{12}^{\dag}\eta_{11}\mp \eta^{\dag}_{22}\eta_{21}),\nn\\[2mm]
{\cal F}_2=&e^{A-C}(|\eta_{11}|^2-|\eta_{12}|^2\mp(|\eta_{21}|^2-|\eta_{22}|^2)),\nn\\[2mm]
{\cal V}=&e^{A}\bigg[\eta_{11}^{\dag}\gamma_{a}\eta_{11}+\eta_{12}^{\dag}\gamma_{a}\eta_{12}\mp(\eta_{21}^{\dag}\gamma_{a}\eta_{21}+\eta_{22}^{\dag}\gamma_{a}\eta_{22})\bigg]e^a\label{eq:ffvdef},\\[2mm]
{\cal U}_i=&2 e^A\big(\text{Re}(\eta_{11}^{c\dag}\gamma_a\eta_{12}\mp\eta_{21}^{c\dag}\gamma_a\eta_{22})e^a~,\text{Im}(\eta_{11}^{c\dag}\gamma_a\eta_{12}\mp\eta_{21}^{c\dag}\gamma_a\eta_{22})e^a~,\text{Im}(\eta_{11}^{\dag}\gamma_a \eta_{12}\mp\eta_{21}^{\dag}\gamma_a \eta_{22})e^a\big)_i,
\end{align}
where $a$ is a flat index on M$_5$. Equipped with these definitions we find for our ansatz that \eqref{eq:extendedKillingvectors} becomes
\begin{align}
K^{IJ}&=2\bigg[-y_i{\cal V}^a\partial_{a} +{\cal F}_1(dy_i)^{\mu}\partial_{\mu}+ {\cal F}_2(K_i)^{\mu}\partial_{\mu}\bigg] (\frac{i}{2}\Sigma^i)^{IJ}+ 2 {\cal U}^a_i\partial_{a} (\frac{i}{2} \tilde{\Sigma}^i)^{IJ},\label{eq:killingvectors}
\end{align}
where $\tilde{\Sigma}^i=(\sigma_2\otimes \mathbb{I},~\sigma_1\otimes \sigma_2,~\sigma_3\otimes \sigma_2)^i$ is another SU(2) representation.
Clearly that this object should be Killing imposes some constraints, as $(K_i)^{\mu}$ are the Killing vectors on the 2-sphere one might imagine that we should arrange for all the rest of the terms to vanish. However, the ansatz we take is consistent with S$^2$ becoming enhanced to a round 3-sphere as
\beq
e^{2C}ds^2(\text{S}^2)+ds^2(\text{M}_5)= e^{2C}\big(d\theta^2+\sin^2\theta ds^2(\text{S}^2)\big)+ ds^2(\tilde{\text{M}}_4).
\eeq
When this is the case both $(K_i)^{\mu}\partial_{\mu}$ and
\beq
\tilde{K}_i=y_i\partial_{\theta}+\cot\theta(dy_i)^{\mu}\partial_{\mu},
\eeq
are SU(2) Killing vectors of respectively  the anti-diagonal and diagonal SU(2) subgroups of  SO(4) $=$ SU(2)$_L\times$SU(2)$_R$. As such when S$^2\to$ S$^3$, the R-symmetry Killing vectors are $(K_i)^{\mu}\partial_{\mu}+\tilde{K}_i$. None the less we shall impose
\beq\label{eq:restriction1}
{\cal V}= {\cal F}_1=0,~~~~{\cal F}_2=\text{constant}\neq 0,
\eeq
by doing so we are excluding the possibility of round 3-sphere solutions. The reasons for  doing  this are the following: First solutions with a round S$^3$ can be mapped to those with a round S$^2$ via duality, as we explain in more detail in section \ref{sec:generalityofansatz} - so all AdS$_3\times$S$^3$ solutions can be generated from a full classification of AdS$_3\times$S$^2$  solutions. Second a main benefit of  AdS$_3\times $S$^3$ vacua is  that they can support ${\cal N}=(4,4)$  supersymmetry. However we are blind to this  using the chiral ${\cal N}=(1,0)$ conditions of \eqref{eq:bps1}-\eqref{eq:bps3}. The ${\cal N}=(1,1)$ conditions of \cite{Macpherson:2021lbr} are better suited to addressing this, as such we leave a study of ${\cal N}=(4,4)$ AdS$_3\times $S$^3$  for future work and focus on solutions satisfying \eqref{eq:restriction1}.\\
~\\ 
The final term ${\cal U}^a_i\partial_{a}$ is a bit more subtle, it is not charged under the SU(2) of S$^2$ but it still generically defines a 3 Killing vectors, this time in M$_5$. Let us stress however that ${\cal U}^a_i\partial_{a}$ need not span all isometries in M$_5$ just the number of a priori isometries supersymmetry demands, more may get imposed by the Bianchi identities or one could choose to impose additional isometries in a more general class. When the $d=5$ spinors are not charged under ${\cal U}^a_i\partial_{a}$, the solution still preserves the small ${\cal N}=(4,0)$ algebra. If the 5d spinors are charged under ${\cal U}^a_i\partial_{a}$ then the algebra experiences some enhancement: For example, if these Killing vectors span a 3-sphere then the ansatz becomes consistent with large ${\cal N}=(4,0)$ on S$^2\times$S$^3$. For small $(4,0)$ specifically then we should impose that ${\cal U}^a_i\partial_{a}$ are flavour isometries, but not that they should be zero in general. Instead one should be able to construct several different types of  solution classes, distinguished by the number of a priori isometries M$_5$ necessarily comes equipped with ( ie 0,1,2,3). That these should be uncharged isometries means that one is free to T-dualise on them: This suggests that one can generate much of what is contained in the small ${\cal N}=(4,0)$ classes with a priori isometries from the classes with no such isometries via duality.

\subsection{On the generality of the round S$^2$ ansatz modulo duality}\label{sec:generalityofansatz}
In this section we shall comment on the generality of the ansatz we have made thus far. In particular we shall illustrate how much of what is omitted by the round 2-sphere ansatz we have made can actually be generated with duality starting from a round S$^2$ classification of both IIA and IIB.\\
~~\\
In \eqref{eq:SU2ansatz} we have made the assumption that the SU(2) R-symmetry is realised by a round S$^2$. Generically one could look for solutions where M$_5$ is fibered over the S$^2$ such that SU(2) is preserved. However the SU(2) of the 2-sphere is only lifted to an isometry of this full space if the connection 1-forms that mediate this transform as gauge fields with respect to SU(2) \cite{Legramandi:2020txf}. As we also need spinors transforming in the $\textbf{2}\oplus \overline{\textbf{2}}$ this restricts the additional possibilities to M$_7$ containing a torus fibration over S$^2$ with spinors that are singlets under of the U(1)s of the torus\footnote{To be more precise we mean the  ${\cal N}=(4,0)$ spinors, a geometry might also support additional spinors charged under one or more of the U(1)s, but they would need to be $(0,n)$ spinors that are singlets of SU(2). Thus they are auxiliary to this argument}, ie the metric must decompose as
\beq
ds^2(\text{M}_7)= \frac{1}{4}\bigg[\sum_{q=1}^ne^{2C_q}D\psi_q^2+e^{2C}ds^2(\text{S}^2)\bigg]+ ds^2(\text{M}_{5-n}),~~~D\psi_q=(d\psi_q- \eta +{\cal A}_q),~~~ d\eta=\text{vol}(\text{S}^2),
\eeq  
where $(e^{A},e^{C_{q}},e^{C},\Phi,{\cal A}_q)$ depend on M$_{5-n}$ only and the fluxes may only depend on the $\mathbb{T}^n$ and S$^2$ directions via $(D\psi_q,\text{vol}(\text{S}^2))$ only. We would now like to establish when a solution of this type can be generated from a round 2-sphere solution by duality. For simplicity let us just assume that $n=1$ so that we have simply an SU(2)$\times$U(1) squashed 3-sphere. The case of generic $n$ works analogously, one simply needs to apply T-duality more times. Such solutions may be distinguished by the form the NS 3-form takes, in general this can depend on the SU(2)$\times$U(1) invariants as
\beq
H_3= h_0D\psi\wedge \text{vol}(\text{S}^2)+D\psi\wedge H_2 +e^{2C_2}H_1\wedge \text{vol}(\text{S}^2)+ \tilde{H}_3 
\eeq
where $h_0$ is a constant. Since $\partial_{\psi}$ should be a flavour isometry one is free to T-dualise on it without breaking the $(4,0)$ supersymmetry. When $h_0 =0$ this maps a solution in IIA/IIB to IIB/IIA with a round 2-sphere. When $h_0\neq 0$ T-duality maps one between solutions with squashed 3-spheres and non trivial $h_0$. However, in the IIA duality frame of such a solution, we must have that the magnetic RR 2-form decomposes as
\beq
f_2= \tilde{g}_0\text{vol}(\text{S}^2)+ \tilde{g}_1\wedge D\psi+\tilde{g}_2,
\eeq
none of which give rise to $D\psi\wedge \text{vol}(\text{S}^2)$ under $d$. Thus similar to our discussion below \eqref{eq:7dbianchis}, the Bianchi identity of the RR 2-form imposes
\beq
f_0 h_0=0.
\eeq  
If we take $f_0=0$ we can lift this solution to $d=11$. Generically such a solution would have a $\mathbb{T}^2$ (spanned by $\psi$ and the M-theory circle)  fibered over S$^2$, but if $(\tilde{g}_0=0,\tilde{g}_1=0)$  the M-theory circle is not fibered over the S$^2$, so if we reduce back to IIA now on $\psi$ we arrive at a round S$^2$ in IIA.  Note in particular that this means that all solutions with a round S$^3$ can be mapped by duality to solutions in type II with a round S$^2$ justifying our assumption \eqref{eq:restriction1}.  More generally one can cover almost all small ${\cal N}=(4,0)$ solutions of type II supergravity modulo duality, by considering  a round S$^2$ in each of IIA and IIB, what remains is rather constrained and could be studied modulo duality from an 11d perspective.\\
~\\
In the next section we shall classify AdS$_3\times$S$^2$ solutions containing no a priori isometry in M$_5$.

\section{Classes of solution with no a priori isometry}\label{eq:noisometry}
In this section we classify solutions on AdS$_3\times$S$^2\times$ M$_5$ that preserve small ${\cal N}=(4,0)$ supersymmetry and contain no a priori isometry in M$_5$. As we shall see, only in IIB does this lead to new classes of solution, specifically 2 distinguished by their compatibility with D7 brane (like) sources.  Class I, for which $dF_1=0$ globally, can be found in section \ref{sec:classI} while class II, for which $dF_1=0$ need only hold away from the loci of sources, can be found in section \ref{sec:classII}. For both classes we reduce the conditions that define the existence of supersymmetric solutions to local expressions for the supergravity fields and a number of PDEs.  We illiterate the methods we employ to achieve this by deriving a  sub-class of class I in explicit detail in section \ref{sec:classI0}.  The general classes are rather broad, so we also consider some restricted cases of interest in sections \ref{eq:S3class} and \ref{eq:classIIrestriction}.\\
~\\
In the previous sections we give criteria to classify solutions on AdS$_3\times$S$^2\times \text{M}_5$, namely we define a set of Killing vector bi-linears \eqref{eq:killingvectors} which can be used to establish when
\begin{enumerate}
\item S$^2$ does not experience an enhancement to S$^3$: ${\cal V}={\cal F}_1=d{\cal F}_2=0$
\item M$_5$ contains additional a priori isometries: spanned by independent components of ${\cal U}_i$ 
\end{enumerate}
We reiterate that we discard 3-sphere solutions as they can all be mapped to round 2-sphere solutions with duality. A class of solutions preserves small ${\cal N}=(4,0)$ rather than large ${\cal N}=(4,0)$ or some more supersymmetric algebra when ${\cal U}_i$ is spanned by flavour isometries in M$_5$. This is  a little tricky to impose on our spinor ansatz a priori, so a practical approach is to classify solutions with 0,1,2,3 a priori isometries one at a time. When we simply fix ${\cal U}_i=0$, there are no a priori isometries in M$_5$  to worry about so solutions certainly preserve just small ${\cal N}=(4,0)$. The focus of the rest of this work will be to classify all such solutions, leaving ${\cal U}_i\neq 0$ for future work. We expect that much of what can be derived for the cases of $1,2,3$ flavour isometries can be generated from classes with no a priori isometry via duality, though we are not claiming that it all can: For instance the round S$^2$ element of the duality orbit of some 3-sphere classes may necessitate 1 a priori isometry. None the less classifying ${\cal U}_i=0$ is a sensible place to start as, mod duality, these are likely the most general classes with exactly\footnote{More correctly at most 8 left chiral supercharges, there could be additional right chiral supercharges.}  small ${\cal N}=(4,0)$.\\
~~\\
In both IIA and IIB it is possible to solve both the 2-sphere and no a priori isometry constraints in general. For IIA, we already know that the SU(2)-structure classes of \cite{Lozano:2019emq}, contain no a prior isometry in M$_5$. Under the assumption that the functions appearing in \eqref{eq:genspinors} are not tuned so as to necessarily reduce the ansatz to SU(2)-structure, it is possible to show that \eqref{eq:zeroformsIIA},\eqref{eq:restriction1} and ${\cal U}_i=0$ can be solved  as
\begin{align}
a&=d_1=1,~~~Z_1=Z_2,~~~~b=d_2=0,~~~~\alpha_2=\frac{1}{2}(\beta_1+\beta_2)=\frac{\pi}{2},\nn\\[2mm]
c&= \frac{\sin\alpha_1\sin\beta_2}{\sqrt{\cos\beta_2^2+\sin^2\alpha_1\sin^2\beta_2}},~~~~d(e^{A-C}\sin\alpha_1)=0
\end{align}
and $\sin\alpha_1\neq 0$ without loss of generality. Unfortunately this ansatz dies off very quickly once the remaining supersymmetry conditions are considered. This can easily be seen from the 1-form part of \eqref{eq:IIfirstbispinorconds3} which demands that
\beq
c\cos\alpha_1=0
\eeq
for this tuning of the functions of the spinor ansatz. Setting either factor in the above to zero reduces the spinors to an SU(2)-structure ansatz anyway. Thus under our assumptions we can conclude that 
\beq
\text{IIA}+ \text{no apriori isometry in M}_5~~~ \Rightarrow~~~ \text{SU(2)-structure}
\eeq
as these solutions are already classified we shall not comment on them further here.\\
~~\\
For our purposes the status of IIB is more promising: It is possible to show that \eqref{eq:zeroformsIIB},\eqref{eq:restriction1} and ${\cal U}_i=0$ can be solved without loss of generality as
\beq\label{eq:IIBtuning}
a=d_1=c=1,~~~Z_1=Z_2=Z,~~~~b=d_2=0,~~~\alpha_1=\frac{\pi}{2}~~~\beta_2=-\beta_1=\beta,~~~~   d(e^{A-C}\cos\alpha)=0
\eeq
where $\cos\alpha_2=\cos\alpha\neq 0$ - note we are dropping the indices on $(\alpha_2,\beta_1)$. Unlike IIA this tuning of the spinor ansatz  necessarily yields an identity-structure, and does not collapse once the rest of the supersymmetry constraints are considered. Plugging the spinors \eqref{eq:genspinors} for the tuning  \eqref{eq:IIBtuning} into \eqref{5deqIIB1}- \eqref{thirdgenIIB}  one finds that these conditions truncate rather a lot, to show this we find it convenient to redefine the vielbein on M$_5$ in terms of $\{U,e^1,e^2,e^3,K\}$ as
\begin{align}
\text{Re}Z&= -\cos\beta K+\sin\beta U,~~~V=\cos\beta U+\sin\beta K,\nn\\[2mm]
e^1&=-\frac{1}{2}\text{Im}\iota_{\overline{Z}}\Omega,~~~e^2=\frac{1}{2}\text{Re}\iota_{\overline{Z}}\Omega,~~~e^3= \text{Im}Z,
\end{align}
with orientation $\text{vol}(\text{M}_7)= e^{2C}\text{vol}(\text{S}^2)\wedge \text{vol}(\text{M}_5)$ for $\text{vol}(\text{M}_5)=U\wedge e^{123}\wedge K$. One then finds that the IIB avatars of \eqref{eq:bps1}-\eqref{eq:bps2} are implied by the following conditions independent of the RR fluxes
\begin{subequations}
\begin{align}
&e^{C}=\frac{1}{2m}e^A \cos\alpha,~~~~ 2me^{2C}H_1=-d(e^{A+C}\cos\beta\sin\alpha)- e^{A}(\cos\beta U-\cos\alpha \sin\beta K),\label{eq:bpsIIBnoisometry1}\\[2mm]
%&d(e^{-A-C+\Phi}\sin\alpha\sin\beta)-e^{-A-2C+\Phi}\sin^2\alpha\sin\beta U=0,\label{eq:bpsIIBnoisometry1.1}\\[2mm]
& d( e^{A}(\cos\beta U-\cos\alpha \sin\beta K)=d(e^{2A-\Phi}\cos\alpha U\wedge K)=0,\label{eq:bpsIIBnoisometry2}\\[2mm]
&d(e^{-A-C+\Phi}\sin\beta \sin\alpha)- e^{-A-2C+\Phi}\sin^2\alpha \sin\beta U= d(e^{2A+C-\Phi} e^i)+e^{2A-\Phi}\sin\alpha U\wedge e^i=0,\label{eq:bpsIIBnoisometry3}\\[2mm]
&\frac{1}{2}\epsilon_{ijk}\bigg[ d\big(e^{2A+C-\Phi}e^j\wedge e^k\wedge (\cos\alpha\sin\beta U+\cos\beta V)\big)+e^{2A-\Phi}\cos\beta \sin\alpha e^j\wedge e^k\wedge U\wedge K\bigg]= e^{2A+C-\Phi}H_3\wedge e^i,\nn\\[2mm]
& d\big(e^{2A-\Phi}e^{123}\wedge (\cos\alpha\cos\beta U- \sin\beta K)\big)= e^{2A-\Phi}\cos\alpha U\wedge K\wedge H_3\label{eq:bpsIIBnoisometry4},
\end{align}
\end{subequations}
and define the components of the magnetic portions of the RR fluxes that are respectively orthogonal and parallel to $\text{vol}(\text{S}^2)$ as
\begin{subequations}
\begin{align}
e^{3A+2C}\star_5 \lambda g_{1+}&= d(e^{3A+2C-\Phi}(\sin\beta U+\cos\alpha\cos\beta K))\label{eq:flux1}\\[2mm]
&-e^{3A+2C-\Phi}\sin\alpha H_1\wedge K+2m e^{2A+2C-\Phi}\cos\beta \sin\alpha U\wedge K\nn\\[2mm]
&- d(e^{3A+2C-\Phi}\cos\alpha e^{123})+e^{3A+2C-\Phi}(\sin\beta U + \cos\alpha \cos\beta K)\wedge H_3\nn\\[2mm]
&+ e^{3A+2C-\Phi}\cos\beta \sin\alpha H_1\wedge e^{123}+2 m e^{2A+2C-\Phi}\sin\alpha e^{123}\wedge U,\nn\\[2mm]
e^{3A}\star_5 \lambda g_{2+}&=- d(e^{3A-\Phi}\sin\alpha K)+2m e^{2A-\Phi}\cos\alpha U\wedge K- e^{3A-\Phi}\sin\alpha K\wedge H_3\label{eq:flux2}\\[2mm]
&+d(e^{3A-\Phi}\cos\beta \sin\alpha e^{123})+2 m e^{2A-\Phi}e^{123}\wedge (\cos\alpha \cos\beta U-\sin\beta K)\nn.
\end{align}
\end{subequations}
Given these, one can eliminate the  flux terms from \eqref{eq:bps3} arriving at a single condition\footnote{generically \eqref{eq:bps3} decomposes in terms of SU(2) singlet and triplet contributions, but in this case the latter are implied} 
\begin{align}
&\bigg(d(e^{2A}\cos\beta \cos\alpha \sin\alpha)\wedge (\sin\beta U+\cos\beta \cos\alpha K)\label{eq:pairingreduced}\\[2mm]
&- \sin(2\alpha)(d(e^{2A}\cos\alpha)+m e^{A}\sin\alpha U)\wedge K\bigg)\wedge e^{123}=0.\nn
\end{align}
Of this system of sufficient conditions for supersymmetry, \eqref{eq:bpsIIBnoisometry1} and \eqref{eq:flux1}-\eqref{eq:flux2} just act as definitions for certain physical fields - it is \eqref{eq:bpsIIBnoisometry2}-\eqref{eq:bpsIIBnoisometry4} and \eqref{eq:pairingreduced} that we must actively solve. To this end one  first needs to define local coordinates and a vielbein from the conditions in \eqref{eq:bpsIIBnoisometry2}-\eqref{eq:bpsIIBnoisometry3}, then plug this into the remaining conditions, extracting $H_3$ and PDEs that imply the remaining conditions. Of course this only fixes the local form of the physical fields up to some PDEs that imply supersymmetry, in addition to solving these, to actually have a solution we must also solve the Bianchi identities of the NS and magnetic RR fluxes which imply additional PDEs. The conditions  \eqref{eq:bpsIIBnoisometry1}-\eqref{eq:pairingreduced} actually give rise to 2 physically distinct classes; $\sin\beta=0$ or not. To establish this one must solve the supersymmetry conditions under the assumptions that $\sin\beta\neq 0$ and $\cos\beta\neq 0$, one finds that the Bianchi identity of $F_1$ is implied by supersymmetry in the former case but not the latter - ie source D7 branes are only possible when $\sin\beta=0$. A special limit is when $\cos\beta=0$, though this is part of the more general $\sin\beta \neq 0$ class, in this limit the PDEs governing the system give rise to a harmonic function constraint common to partially localised brane intersections \cite{Youm:1999ti}. \\
~\\
In the next section we give the local form of the class of solution with $\cos\beta =0$ turning our attention to the general case in section \ref{sec:classI}.
%%%%%%%%%%%%%%%%%%%%%%%%%%%%%%%%%%%%%%%%%%%%%%%%%%%%%%%%%%%%%%%%%%%%%%%%%%
%%%%%%%%%%%%%%%%%%%%%%%%%%%%%%%%%%%%%%%%%%%%%%%%%%%%%%%%%%%%%%%%%%%%%%%%%%
%%%%%%%%%%%%%%%%%%%%%%%%%%%%%%%%%%%%%%%%%%%%%%%%%%%%%%%%%%%%%%%%%%%%%%%%%%
\subsection{$\cos\beta =0$ : A sub-class of class I with harmonic function rule}\label{sec:classI0}
In this section we derive the class of solutions that follows from fixing $\cos\beta=0$, we can set $\beta=\frac{\pi}{2}$ without loss of generality. This is actually a limit of the more general class I with $\sin\beta \neq 0$ in the next section but classifying that in this limit has value as it gives is relatively simple class, which the general classes are not. First it serves as a warm up illustrating how we reduce the supersymmetry constraints and Bianchi identities to physical fields and PDEs -  in the more general classes we use the same methods but  give a less detailed derivation. Second it gives some understanding of the types of solutions that class I  contains.\\
~~\\
We begin by noting that the second of \eqref{eq:bpsIIBnoisometry3} implies $d(e^{-C}\sin\alpha U)=0$. One might be tempted to use this to define a local coordinate, however this would not be valid when $\sin\alpha=0$. In general it implies that we can introduce a function $f$ on M$_5$ and then integrate the second of \eqref{eq:bpsIIBnoisometry3} as follows
\beq
 e^{-C}\sin\alpha U=d\log f~~~~\Rightarrow    e^{2A+C-\Phi}f e^i = dz_i
\eeq
where $z_i$ for $i=1,2,3$ are local coordinates on M$_5$, which fixes 3 components of the vielbein. We define the final 2 components with \eqref{eq:bpsIIBnoisometry2}: We integrate the first of these in terms of a local coordinate $y$ as
\beq
e^A \cos\alpha K=-dy
\eeq
so that the second condition becomes $d(e^{A-\Phi}U)\wedge dy$. One can show that this can be integrated in general in terms of a final local coordinate $x$ and $\lambda=\lambda(x,y,z_i)$ as
\beq
e^{A-\Phi}U= dx+ \lambda dy,
\eeq
at which point a set of local coordinates and vielbein on M$_5$ are determined without loss of generality. Plugging this back into \eqref{eq:bpsIIBnoisometry3} we find first that $P=P(x,y)$, and then
\beq 
2m\tan\alpha=e^{2A-\Phi}\partial_{x}\log f,~~~~d(f^{-2}\partial_xf)=0,~~~~\partial_y f =\lambda\partial_x f.
\eeq
In general these imply $cf^{-1}= x+ g(y)$ for $c$ a constant, and $\lambda=\partial_y g$ - however we then have that $U\sim d(x+g)$, so up to a change in coordinates we can simply fix
\beq\label{eq:caseI0bps1} 
\lambda=0,~~~~~  f^{-1}= u(x),~~~~ u''=0,
\eeq
making the metric diagonal. Next we note that  the first of \eqref{eq:bpsIIBnoisometry4} wedged with $v_i$ is independent of $H_3$, so gives rise to constraints on the local functions derived thus far: This and \eqref{eq:pairingreduced} respectively imply
\beq\label{eq:caseI0bps2}
\partial_y(e^{2A-\Phi})=0,~~~~\partial_x(e^{4A}u^{-1}\cos^2\alpha)=0,
\eeq
which together with \eqref{eq:caseI0bps1} are actually all the PDEs supersymmetry demands that we solve as shall be come clear momentarily. Supersymmetry also demands that we solve \eqref{eq:bpsIIBnoisometry4} so to proceed we take a general ansatz for $H_3$ in terms of our local coordinates and 10 functions with support on M$_5$
\beq
H_3=H^0dz_1\wedge dz_2\wedge dz_3+ H^0_i dx\wedge dy\wedge  dz_i+\frac{1}{2}H^x_i\epsilon_{ijk}dx\wedge dz_j\wedge dz_k+\frac{1}{2}H^y_i\epsilon_{ijk}dy\wedge dz_j\wedge dz_k.
\eeq
Inserting this into \eqref{eq:bpsIIBnoisometry4}, making use of what has been derived thus far, we find that it just implies
\begin{align}
H^0&=8 m^3\partial_x\left(\frac{1}{e^{8A-2\Phi}\cos^4\alpha}\right),\nn\\[2mm]
H^0_i&=H^y_i=0,~~~~H^x_i=-2m\partial_{z_i} (e^{-4A+2\Phi}u).
\end{align}
What remains to be dealt with is \eqref{eq:bpsIIBnoisometry1} and \eqref{eq:flux1}-\eqref{eq:flux2}, which are all just definitions of physical fields. We do however still need to take the hodge dual of the latter to construct the magnetic RR fluxes - this is not difficult as we have an explicit vielbein to work with, we find
\begin{align}
f_1&=f_7=0, ~~~~e^{2C}H_1=-\frac{1}{2m}dy,\nn\\[2mm]
f_3&=-\frac{1}{4m^2}\left(e^{2A-\Phi}\cos\alpha\sin\alpha-2mx\right)\wedge \text{vol}(\text{S}^2)+m\epsilon_{ijk}\partial_{z_i}\left(\frac{u}{e^{4A}\cos^2\alpha}\right)dz_j\wedge dz_k\wedge dy\nn\\[2mm]
&-8m^3 e^{-4A+2\Phi}(u')^2\partial_y(\frac{u}{e^{4A}\cos^2\alpha})dz_1\wedge dz_2\wedge dz_3,\\[2mm]
f_5&=\frac{e^{4A-2\Phi}}{4m^2 \cos^2\alpha u^2}\bigg[ e^{2A-\Phi} u\tan\alpha\epsilon_{ijk}\partial_{z_i}\left(e^{-4A+2\Phi}u^2\right)dz_j\wedge dz_k\wedge dx\nn\\[2mm]
& \frac{u}{e^{4A}\cos^2\alpha}\bigg(16m^4 e^{-4A+2\Phi}u^4- 8m^2 e^{-2A+\Phi}u^3\tan\alpha\partial_x\left(e^{-4A+2\Phi}u\right) \bigg)dz_1\wedge dz_2\wedge dz_3\bigg]\wedge \text{vol}(\text{S}^2)\nn.
\end{align}
We have now reduced the supersymmetry conditions to definitions of the physical fields and 3 PDEs \eqref{eq:caseI0bps1}-\eqref{eq:caseI0bps2}, but we still need to solve the Bianchi identities of the fluxes to have a solution. If we assume only (at least partially) localised sources, away from the loci of these this amounts to solving $dH_3=0$ and $d_Hf_-=0$ which must hold in all regular regions of a solution with or without sources. Our approach to deal with the Bianchi identities will be to assume  we are in a local regular region and reduce the Bianchi identities to a set of PDES that define the class. For specific solutions one then needs to check whether this local region can be extended to potential singular loci, ie one must additionally check  that the  PDEs give appropriate $\delta$-function sources and that these have a supersymmetric embedding. Away from the loci of sources the Bianchi identities of $H_3$ and $f_3$ impose
\begin{align}
&\partial_y\left(\frac{u}{e^{4A}\cos^2\alpha}\right)\partial_x\left(e^{-4A+2\Phi}u^2\right)=0,\nn\\[2mm]
&\partial_{z_i}^2\left(e^{-4A+2\Phi}u^2\right)+ 4m^2\frac{u^2}{e^{4A}\cos^2\alpha} \partial_{x}^2\left(e^{-4A+2\Phi}u^2\right)=0,\nn\\[2mm]
&\partial_{z_i}^2\left(\frac{u}{e^{4A}\cos^2\alpha}\right)+4m^2e^{-4A+2\Phi}u^2 \partial_{y}^2\left(\frac{u}{e^{4A}\cos^2\alpha}\right)=0
\end{align}
The first of these is a harmonic function rule that induces a splitting of the class into 2 cases, depending on which factor vanishes. The Bianchi identity for $f_5$ is implied by these and the supersymmetry conditions. At this point we have reduced the conditions to have a supersymmetric solution to some PDEs - but we can express the class in a more concise fashion in terms of some new functions as
\beq
P=\frac{ 4m^2 u}{e^{4A}\cos^2\alpha},~~~~ G=4m^2e^{-4A+2\Phi}u^2,
\eeq
in terms of which \eqref{eq:caseI0bps2} become simply $G=G(x,z_i)$, $P=P(y,z_i)$.  At this point the class is defined by just the Bianchi identities.  This completes our derivation of the PDEs governing this class.\\
~\\
In summary the class of solutions in this section has an NS sector of the form
\begin{align}
ds^2&=2m\bigg[\sqrt{1+ \frac{(u')^2}{G}}\bigg(\sqrt{\frac{ u}{ P}}ds^2(\text{AdS}_3)+\frac{G}{4m^2 \sqrt{u}}\left(\frac{1}{\sqrt{P}u}dx^2+\sqrt{P}dz_i^2\right)+\frac{1}{4m^2}\sqrt{\frac{P}{u}}dy^2\bigg)\nn\\[2mm]
& +\frac{1}{4m^2\sqrt{1+ \frac{(u')^2}{G}}}\sqrt{\frac{u}{P}}ds^2(\text{S}^2)\bigg],~~~~~e^{-\Phi}=\sqrt{\frac{P u}{G\left(1+\frac{(u')^2}{G}\right)}},\nn\\[2mm]
2mH&=-dy\wedge \text{vol}(\text{S}^2)-\frac{ 1}{2 u} \epsilon_{ijk}\partial_{z_i}Gdx\wedge dz_j\wedge dz_k+P\partial_x G dz_{123},
\end{align}
where $G=G(x,z_i)$,  $P=P(y,z_i)$, $u$ is a linear function of $x$ and we use the short hand notation $dz_{123}= dz_1\wedge dz_2\wedge dz_3$. The non trivial ten dimensional RR fluxes are 
\begin{align}
F_3&=\frac{1}{2m}d\left(\frac{uu'}{G+(u')^2}-x\right)\wedge \text{vol}(\text{S}^2)+\frac{1}{4m}\epsilon_{ijk}\partial_{z_i}Pdy\wedge dz_j\wedge dz_k-\frac{1}{2m}G\partial_y P dz_{123},\\[2mm]
F_5&=(1+\star_{10})f_5,\nn\\[2mm]
f_5&=\frac{1}{8m^3(G+(u')^2)}\bigg[mu' \epsilon_{ijk}\partial_{z_i}Gdx\wedge dz_j\wedge dz_k+2mP\bigg(G^2- u' u^2 \partial_x(Gu^{-1})\bigg)dz_{123}\bigg]\wedge \text{vol}(\text{S}^2)\nn.
\end{align}
One has a solution whenever the Bianchi identities of the fluxes are satisfied, away from possible sources these impose
\beq
\partial_xG \partial_y P=0,~~~~\partial_{z_i}^2P+G\partial_y^2P=0,~~~~\partial_{z_i}^2G+u P\partial_x^2G=0.
\eeq
More generically the latter two of these could have $\delta$-function sources, when this is the case they should have a supersymmetric embedding for the remaining equations of motion of a solution to necessarily follow. The first condition is a harmonic function rule: It states that either $\partial_x G=0$, or $\partial_y P=0$ yielding two cases.\\
~~\\
\textbf{Case 1}\\
~\\
For the first case, when $\partial_x G=0$, the Bianchi identities reduce to those of D5 branes ending on  NS5 branes smeared over 1 direction in flat space \cite{Youm:1999ti} (see section 4.5). Comparing to this suggests  that this case formally describes localised D5 branes of world volume $(\text{AdS}_3,\text{S}^2,x)$ ending at NS5 branes on $(\text{AdS}_3,\text{S}^2,y)$ that are delocalised in $x$. Unlike the flat space case,  $\partial_x$ is not an isometry of the solution in general, the warping in $G$ is more complicated and additional fluxes and flux components are turned on. The latter difference is to be expected as Mink$_5$ is replaced by AdS$_3\times$S$^2$. Though the warping in $G$ is more complicated than the flat space case, notice that as $G$ becomes large it does tend to what one would expect in the flat space case. The dependence of the metric on $u(x)$  is essentially a deformation of this system: when $u=1$, the warp factors becomes precisely what one would expect for this D5-NS5 system and $x$ becomes an isometry, so we can take it to be a compact direction. When $u \neq 1$, as nothing else depends on $x$ this direction is unbounded.\\
~~\\
\textbf{Case 2}\\
~\\
For the second case when $\partial_y P=0$ the Bianchi identities do not reduce to those of a simple flat space brane intersection, though they are not far removed - the issue is the function $u$. As $P$ is the only object with $y$ dependence generically, this now becomes an isometry for this case. Examining the metric and NS and RR fluxes, it should not be hard to see that performing T-duality on $\partial_y$ maps to a class of solutions in IIA with a round 3-sphere (locally) and non trivial fluxes $(H,F_2,F_4)$. In IIB when $u=1$ the Bianchi identities reduce to what one would expect for localised NS5 branes of world volume $(\text{AdS}_3,\text{S}^2,y)$ ending on  D5 brane of world volume (AdS$_3$, S$^2$, $x$) that are smeared over $y$, when $u\neq 1$ we have a deformation of this system. Even when $u\neq 1$ for large $G$ the functions $P,G$ appear where one would expect for D5 and NS5 brane warp factors respectively, but the additional $u$ dependence in the metric further deforms this picture. As we shall explain at greater length in section \ref{eq:S3class} this case represents a  of a class derived in \cite{Faedo:2020nol}, which is actually related to an SU(2)-structure class in IIA via duality. \\
~~\\
Having derived and interpreted the class with $\cos\beta=0$, we shall now move onto its generalisation with $\sin\beta \neq 0$ in the next section.
%%%%%%%%%%%%%%%%%%%%%%%%%%%%%%%%%%%%%%%%%%%%%%%%%%%%%%%%%%%%%%%%%%%%%%%%%%%%%%%%%%%%%%%%%%%%%%%%%%%%%%%%%%%%%%%%%%%%%%%%%%%%%%%%%%%%%%%%%%%%%%%%%%%%%%%%%%%%%%%%%%%%%%%%%%%%%%%%%%%%%%%%%%%%%%%%%%%%%%%%%%%%%%%%%%%%%%%%%%%%%%%%%%%%%%%%%%%%%%%%%%%%%%%%%%%%%%%%%%%%%%%%%%%%%%%%%%%%%%%%%%%%%%%%%%%%%%%%%%%%%%%%%%%%%%%%%%%%%%%%%%%%%%%%%%%%%%%%%%%%%%%%%%%%%%%%%%%%%%%%%%%%%%%%%%%%%%%%%%%%%%%%%%%%%%%%%%%%%%%%%%%%%%%%%%%%%%%%%%%%%%%%%%%%%%%%%%%%%%%%%%%%%%%%%%%%%%%%%%%%%%%%%%%%%%%%%%%%%%%%%%%%%%%%%%%%%%%%%%%%%%%%%%%%%%%%%%%%%%
\subsection{$\sin\beta \neq 0$ : Class I}\label{sec:classI}
The class of the previous section is actually a sub-case of the more general class we consider in this section, this is consistent with fixing $\cos\beta=0$ but does not require it. The method of reducing this class to PDEs is analogous to that of the previous section, so we will be more brief.\\
~~\\
We again use  \eqref{eq:bpsIIBnoisometry2}-\eqref{eq:bpsIIBnoisometry3} to define the vielbein and dilaton in terms of local coordinates $(x,y,z_i)$, this time as
\begin{align}
e^{A}(\cos\beta U-\cos\alpha \sin\beta K)&=dy,~~~~\frac{e^{A-\Phi}\cos\alpha}{\cos^2\beta+\cos^2\alpha\sin^2\beta}(\cos\alpha\sin\beta U+\cos\beta K)= dx+ \lambda dy,\nn\\[2mm]
e^{2A+C-\Phi}f e^i&=dz_i,
\end{align}
without loss of generality - note that this makes the metric on M$_5$ non diagonal generically becoming diagonal for $\cos\beta=0$. Plugging this ansatz into what remains non trivial in \eqref{eq:bpsIIBnoisometry2}-\eqref{eq:bpsIIBnoisometry3}, fixes $\lambda$ to a value we shall quote momentarily and up to diffeomorphisms imposes
\beq
2m\tan\alpha\sin\beta=- e^{2A-\Phi}\partial_x \log u(x),~~~~~ u''=0,
\eeq
which we can take to define $\alpha$. Now as it simplifies the final result we find it convenient to introduce functions $(G,P,Q)$ with support on M$_5$ as
\begin{align}
GP\Delta_2\cot^2\beta&=Q^2u,~~~~ e^{4A}P\Delta_1= 4m^2 u \left(1+\frac{(u')^2}{G}\right),~~~~ e^{4A-2\Phi}G\Delta_1\left(1+\frac{(u')^2}{G}\right)=4m^2 u \Delta_2,\nn\\[2mm]
\Delta_1 &= 1- \frac{u Q^2}{GP},~~~~ \Delta_2= \Delta_1+ \frac{(u')^2
}{G}.
\end{align}
With these definitions the expression for $\lambda$ that follows from \eqref{eq:bpsIIBnoisometry2}-\eqref{eq:bpsIIBnoisometry3} is simply 
\beq
\lambda= \frac{Qu}{G}.
\eeq
Thus when $Q=0$ the class reduces to  the previous section. We then essentially follow the same steps as in the previous section to reduce  \eqref{eq:bpsIIBnoisometry1}-\eqref{eq:pairingreduced} to physical fields and PDEs. The computation is of course more demanding but in essence the only real difference is that supersymmetry demands that the following PDEs are satisfied 
\beq\label{eq:caseIbps}
\partial_yQ= \partial_x P,~~~~ \partial_y G= u \partial_x Q,
\eeq
which are less trivial than those of the previous section. There the PDEs could simply be integrated while now the possible geometries depend on how \eqref{eq:caseIbps}  are solved. Note that \eqref{eq:caseIbps} gives a definition for $\partial_y\partial_x Q$ and $\partial_x\partial_y Q$ whose consistency implies\\
\beq
\partial_y^2 G= u \partial_x^2 P,
\eeq
at least for a sufficiently smooth $Q$. In addition to this we find that the NS sector must take the form
\begin{align}
ds^2&=2m\sqrt{1+\frac{(u')^2}{G}}\bigg[\sqrt{\frac{u}{ P \Delta_1}}\bigg(ds^2(\text{AdS}_3)+\frac{1}{4m^2}\frac{\Delta_1}{\Delta_2}ds^2(\text{S}^2)\bigg)+ \frac{1}{4m^2}\bigg(\sqrt{\frac{P}{u\Delta_1}}Dy^2\nn\\[2mm]
&+G\sqrt{\frac{\Delta_1}{u}}\left(\frac{1}{\sqrt{P}u}dx^2+\sqrt{P}dz_i^2\right)\bigg)\bigg],~~~~e^{-\Phi}=\frac{\sqrt{\frac{P u}{G}}\sqrt{\Delta_2}}{1+ \frac{(u')^2}{G}}\label{eq:classINS}\\[2mm]
2mH&= -d\left(\frac{Q u u'}{GP \Delta_2}+y\right)\wedge \text{vol}(\text{S}^2)-\frac{1}{2}\epsilon_{ijk}\left(\frac{1}{u}\partial_{z_i}Gdx+\partial_{z_i}Qdy\right)\wedge dz_j\wedge dz_k+\partial_x(GP\Delta_1)dz_{123},\nn
\end{align}
where we define the following to make the expressions more compact
\beq
\Delta_1 = 1- \frac{u Q^2}{GP},~~~~ \Delta_2= \Delta_1+ \frac{(u')^2
}{G},~~~~ Dy= dy +\frac{Q}{P}dx,~~~~dz_{123}=dz_1\wedge dz_2\wedge dz_3
\label{eq:def}
\eeq
clearly the metric on M$_5$ contains cross terms in $(x,y)$ generically - we remind the reader that $u$ is a linear function of $x$. In addition to this, the class has all possible ten dimensional RR fluxes non trivial, they take the form
\begin{align}
F_1&= dC_0,~~~~ C_0=-\frac{ Q u}{(G+ (u')^2)},~~~~F_5= (1+\star_{10})f_5\label{eq:classIRR}\\[2mm]
F_3&=-C_0 H +\frac{1}{2m}\bigg[d\left(\frac{uu'}{G\Delta_2}-x\right)\wedge \text{vol}(\text{S}^2)+\frac{1}{2}\epsilon_{ijk}\left(\partial_{z_i}Qdx+\partial_{z_i}Pdy\right)\wedge dz_j\wedge dz_k- \partial_y(GP\Delta_1)dz_{123}\bigg],\nn\\[2mm]
f_5&=\frac{P}{4m^2 G \Delta_2}\bigg[\frac{u'
}{2}\epsilon_{ijk}\bigg(u \partial_{z_i}(QP^{-1})dy+\left(P^{-1}\partial_{z_i}G-\frac{u}{2P^2}\partial_{z_i}(Q^2)\right)dx\bigg)\wedge dz_j\wedge dz_k\nn\\[2mm]
&+P^{-2}\bigg(G^2P^2\Delta_1^2+u u' \left(G^{-2}\partial_y(G^3P Q)-P^2\partial_xG-2u Q^2\partial_x P-u G\partial_{x}(u^{-1}P^2)\right)\bigg)dz_{123}\bigg]\wedge \text{vol}(\text{S}^2)\nn
\end{align}
The definitions of the physically fields in \eqref{eq:classINS}-\eqref{eq:classIRR} along with the PDEs of \eqref{eq:caseIbps} imply supersymmetry. Again to have a solution we need to solve the Bianchi identities of the fluxes,   although the fluxes are a little complicated their Bianchi identities are not especially as many components are implied by \eqref{eq:caseIbps}. First off clearly $dF_1=0$ is implied, so we can have no D7 brane sources in this class. As with the previous case the Bianchi identities of $(H,F_3)$ imply that of $F_5$, in regular regions of a solution these impose the following
\begin{align}\label{eq:classIbianchi}
\partial_{z_i}^2Q+\partial_{x}\partial_y(GP\Delta_1)=0,~~~~\partial_{z_i}^2G+u\partial^2_{x}(GP\Delta_1)=0,~~~~\partial_{z_i}^2P+\partial^2_{y}(GP\Delta_1)=0,
\end{align}
which are clearly more exotic generically than the PDEs one would expect of a simple brane intersection, but are still reminiscent of this. Solutions in this class are in 1 to 1 correspondence with the solutions to the combined systems of \eqref{eq:caseIbps} and \eqref{eq:classIbianchi}.\\
~~\\
In the next section we shall derive a class of solutions in type IIA with a squashed and fibred 3-sphere that follows from imposing that $\partial_y$ is an isometry direction and then T-dualising on it.

\subsubsection{A IIA class with fibered 3-sphere}\label{eq:S3class}
In this section we derive a new class of solutions in IIA via T-duality from class I. Clearly there is generically no isometry to perform this duality on, so we must impose one on the class - we shall take $\partial_y$ to be an isometry. This means that the metric \eqref{eq:classINS} should be $y$ independent - examining the various metric components and given \eqref{eq:caseIbps} this reduces the  conditions for a solution to
\beq
P=P(z_i),~~~~Q=Q(z_i),~~~~ G=G(x,z_i),~~~\partial_{z_i}^2P=0,~~~\partial_{z_i}^2Q=0,~~~\partial_{z_i}^2G+u P\partial^2_{x}G=0
\eeq
Performing T-duality on the $\partial_y$ direction in \eqref{eq:classINS}-\eqref{eq:classIRR} then results in the following NS sector
\begin{align}
\frac{ds^2}{2m}&=\sqrt{1+\frac{(u')^2}{G}}\bigg[\sqrt{\frac{u}{ P \Delta_1}}ds^2(\text{AdS}_3)+ \frac{1}{4m^2}G\sqrt{\frac{\Delta_1}{u}}\left(\frac{1}{\sqrt{P}u}dx^2+\sqrt{P}dz_i^2\right)\bigg]\nn\\[2mm]
&+\frac{1}{4m^2}\frac{1}{\sqrt{1+\frac{(u')^2}{G}}}\sqrt{\frac{u\Delta_1}{P}}\bigg(D\phi^2+\frac{G+(u')^2}{G\Delta_2}ds^2(\text{S}^2)\bigg)\nn\\[2mm]
e^{2\Phi}&=\frac{2mG}{\Delta_2}\sqrt{\frac{\Delta_1}{uP^3}\bigg(1+\frac{(u')^2}{G}\bigg)^3},~~~
B_2=B-\frac{1}{2m}\eta\wedge d\left(\frac{Quu'}{GP\Delta_2}\right)+\frac{Q}{2mP}dx\wedge D\phi,
\end{align}
where we introduce 1-forms $(D\phi,\eta,{\cal A})$ and 2-form $B$ such that
\begin{align}
D\phi&=d\phi+\mathcal{A}+\eta,\qquad \text{with}\qquad d\mathcal{A}=-\frac{1}{2}\epsilon_{ijk}\partial_{z_i}	Qdz_j\wedge dz_k,\qquad
d\eta=\text{vol}(\text{S}^2),\nn\\[2mm]
dB&=\frac{1}{2m}\bigg(\partial_x(PG\Delta_1)dz_{123}-\frac{1}{2}\epsilon_{ijk}\frac{\partial_{z_i}G}{u}dx\wedge dz_j\wedge dz_k\bigg),
\end{align}
here $\phi$ is the dual coordinate to $y$  after a rescaling. 
Notice that $(D\phi,\text{S}^2)$ now span an SU(2)$\times$U(1) preserving squashed and fibered 3-sphere. In addition, the background is supported by the RR fluxes
\begin{align}
	F_0&=0,\qquad
	F_2=\frac{1}{4m}\epsilon_{ijk}\left(\partial_{z_i} P+C_0\partial_{z_i}Q\right)dz_j\wedge dz_k+\frac{C_0}{2m}\text{vol}(\text{S}^2)+ \frac{1}{2m}D\phi\wedge dC_0,\nn\\[2mm]
	F_4&=2m\frac{Quu'}{GP\Delta_1}(P\partial_{z_i}Q-Q\partial_{z_i}P)dz_i\wedge\text{vol}(\text{AdS}_3)
	+\frac{Puu'}{8m^2G\Delta_2}\epsilon_{ijk}\;\partial_{z_i}(QP^{-1})dz_j\wedge dz_k\wedge\text{vol}(\text{S}^2)\nn\\[2mm]
	&+\frac{1}{4m^2}\bigg[C_0\;\partial_x(PG\Delta_1)dz_{123}+\frac{1}{2}\epsilon_{ijk}\left(C_0\bigg(\frac{Q}{P}\partial_{z_i}Q-\frac{\partial_{z_i}G}{u}\bigg)-\big(\partial_{z_i}Q
	+\frac{Q}{P}\partial_{z_i}P\big)\right)dx\wedge dz_j\wedge dz_k\nn\\[2mm]
	& +\bigg(dx-d\bigg(\frac{uu'}{G\Delta_2}\bigg)-C_0\;d\bigg(\frac{Quu'}{QP\Delta_2}\bigg)\bigg)\wedge\text{vol}(\text{S}^2)\bigg]\wedge D\phi
	%&+\bigg(\frac{1}{2mc}\left[dx\big(\partial_x-\frac{Q}{P}\partial_y\big)+dz_i\partial_{z_i}\right]\left(x-\frac{P(\Delta_2-1)}{Q^2\Delta_2}\right)-C_0\left[dx\big(\partial_x-\frac{Q}{P}\partial_y\big)+dz_i\partial_{z_i}\right]\left(\frac{\Delta_2-1}{Q\Delta_2}-y\right)\bigg)\wedge\text{vol}(\text{S}^2)\nn\\
	,
\end{align}
where $C_0, \Delta_1$ and $\Delta_2$ are defined as in \eqref{eq:def}-\eqref{eq:classIRR}.\\
~~\\
To interpret this class it is instructive to first fix $Q=0$ and $u=1$ so that $\Delta_1=1,~ G\Delta_2=G+(u')^2$ and the 3-sphere spanned by $(D\phi,\text{S}^2)$ becomes the round one.  We then find that the governing PDE of the system reduces to $\partial_{z_i}^2G+ P\partial^2_{x}G=0$, the  same PDEs as the system of fully localised NS5 branes inside the world volume of D6 branes derived in \cite{Itzhaki:1998uz}. Here however, rather than Mink$_6$, the branes share the world volume directions AdS$_3\times $S$^3$ with the D6 further extended in $x$. In this limit  $P$ and $G$ appear where one would expect for the warp factors of such  D6 and NS5 branes,  this is also true when $u\neq 1$ but $G$ is large. The effect of turning on $Q$ appears to be to place formal KK monopoles into this D6-NS5 brane system which squashes the 3-sphere. The effect of turning on $u$ is then a deformation.

This case actually generalises an known class of solutions: In \cite{Lozano:2019emq}, there is a class of solution with  D4 branes extended on AdS$_3\times$S$^2$  and localised in CY$_2$ times an interval, that lie inside the world volume of D8 branes. This set up can actually be realised as a near horizon limit of intersecting branes, however the class in \cite{Lozano:2019emq} is a generalisation of this which depends on a linear function of the interval $\tilde{u}$ that cannot be so realised\footnote{Or at least has not yet been so realised. Additionally this class depends in on a primitive $(1,1)$ on CY$_2$ that we assume is set to zero here}. Upon setting the Romans mass to zero this class may be lifted to an  AdS$_3\times $S$^3\times$CY$_2$ class in d=11 with source M5 branes and this additional $\tilde{u}$ deformation (see \cite{Lozano:2020bxo}). If one then imposes that CY$_2= \mathbb{R}^4$ expressed in polar coordinates, and assumes SO(4) rotational symmetry for the M5 brane warp factors, one can reduce back to IIA, but this time on the Hopf fibre of a second 3-sphere $\tilde{\text{S}}^3\subset \mathbb{R}^4$.  Upon fixing $\tilde{u}=1$ the resulting class is  a system of localised D6 branes with localised NS5 branes inside them both extended in AdS$_3\times $S$^3$ under the assumption that their common co-dimensions preserve an SO(3) isometry (see appendix B of \cite{Faedo:2020nol}). Generically the presence of $\tilde{u}$ deforms this system. This sounds quite similar to the $Q=0$ limit of what we have here if we impose an SO(3) isometry in the directions $z_i$. Indeed the SO(3) preserving $Q=0$ limit of the case we present here can be precisely mapped to  the class  of \cite{Faedo:2020nol}.

We have thus constructed a broad generalisation of a class of  AdS$_3\times $S$^3$ solutions found in \cite{Faedo:2020nol}, such that formal KK monopoles are included and the internal space has no SO(3) isometry generically.

%%%%%%%%%%%%%%%%%%%%%%%%%%%%%%%%%%%%%%%%%%%%%%%%%%%%%%%%%%%%%%%%%%%%%%%%%%%%%%%%%%%%%%%%%%%%%%%%%%%%%%%%%%%%%%%%%%%%%%%%%%%%%%%%%%%%%%%%%%%%%%%%%%%%%%%%%%%%%%%%%%%%%%%%%%%%%%%%%%%%%%%%%%%%%%%%%%%%%%%%%%%%%%%%%%%%%%%%%%%%%%%%%%%%%%%%%%%%%%%%%%%%%%%%%%%%%%%%%%%%%%%%%%%%%%%%%%%%%%%%%%%%%%%%%%%%%%%%%%%%%%%%%%%%%%%%%%%%%%%%%%%%%%%%%%%%%%%%%%%%%%%%%%%%%%%%%%%%%%%%%%%%%%%%%%%%%%%%%%%%%%%%%%%%%%%%%%%%%%%%%%%%%%%%%%%%%%%%%%%%%%%%%%%%%%%%%%%%%%%%%%%%%%%%%%%%%%%%%%%%%%%%%%%%%%%%%%%%%%%%%%%%%%%%%%%%%%%%%%%%%%%%%%%%%%%%%%%%%%%%%%%%%%%%%%%%%%%%%%%%%%%%%%%%%%%%%%%%%%%%%%%%%%%%%%%%%%%%%%%%%%%%%%%

\subsection{$\sin\beta = 0$ : Class II}\label{sec:classII}
In this section we derive the class of solutions that follows from fixing $\sin\beta=0$ in the supersymmetry conditions, specifically we will fix $\beta =0$ without loss of generality.\\
~\\
Fixing $\beta=0$ in \eqref{eq:bpsIIBnoisometry2}-\eqref{eq:bpsIIBnoisometry3} changes the character of the solutions some what as we shall see. We can still take the same terms in  \eqref{eq:bpsIIBnoisometry2}-\eqref{eq:bpsIIBnoisometry3} to define the vielbein, this time as
\beq\label{eq:caseIIviel}
e^{A}U= dx,~~~~~ e^{A-\Phi} \cos\alpha K= dy+\tilde{\lambda} dx,~~~~~e^{2A+C-\Phi}f e^i= \frac{1}{2m} dz_i,
\eeq
for $\tilde{\lambda}$ and $f$ functions of all the coordinates on M$_5$.
Again \eqref{eq:bpsIIBnoisometry2}-\eqref{eq:bpsIIBnoisometry3} contain additional constraints which allow us to fix
\beq\label{eq:alphadef}
f=u^{-1},~~~~\tan\alpha=-\frac{1}{2m}e^{2A}\partial_x \log u(x),~~~~ u''=0,
\eeq
without loss of generality. We now find it helpful to introduce functions $h,g,\lambda$ with support on $(x,y,z_i)$ such that
\beq
e^{2A-\Phi}=2m\sqrt{\frac{u\Xi }{g }},~~~~e^{4A}=\frac{g u^2}{h},~~~~\tilde\lambda = \frac{\lambda}{g},~~~~~\Xi= 1+\frac{g(u')^2}{4m^2 h}.
\eeq
This reduces \eqref{eq:bpsIIBnoisometry4} to a single PDE and definition of $H_3$, namely
\beq\label{eq:caseIIbps}
\partial_{y}\lambda=\partial_x g,~~~~~ 4m^2H_3=-\frac{1}{2}\epsilon_{ijk}(\partial_{z_i}g dy+\partial_{z_i}\lambda dx)\wedge dz_j\wedge dz_k+ \partial_yh dz_{123}.
\eeq
What remains of the supersymmetry conditions just defines $H_1$ and the Hodge dual of the magnetic parts of the RR fluxes - the Hodge dual can be taken with respect to \eqref{eq:caseIIviel} without difficulty. Let us just explain how we extract the Bianchi identities before then summarising our results for this class. First \eqref{eq:caseIIbps} clearly implies that away from sources, where $dH_3=0$ should hold we must have
\beq
\partial_{z_i}^2 g+ \partial_{y}^2h=0,~~~~~\partial_{z_i}^2 \lambda+ \partial_{x}\partial_{y}h=0.
\eeq
Moving onto the RR sector, we find that the 1-form is
\beq
f_1=-d\left(\frac{\lambda}{g}\right)+dx \bigg(\partial_x\left(\frac{\lambda}{g}\right)-\frac{1}{2}\partial_y\left(\frac{\lambda^2}{g^2}\right)-\frac{4m^2 u}{g}\partial_y\left(\frac{h}{g^2 u^2}\right)\bigg).
\eeq
As such, away from the loci of sources we must have that the coefficient of $dx$ in this expression is a function of $x$ only for $df_1=0$ to hold.  We can thus introduce a function $w=w(x)$ and identify  $f_1=dC_0$ for
\begin{align}
C_0&=w-\frac{\lambda}{g},
\end{align}
The remaining magnetic RR fluxes can then be more compactly expressed in terms of the following auxiliary functions
\beq
{\cal S}=w\partial_y h-\partial_x h,~~~~{\cal T}=\lambda-w g,~~~~{\cal X}=\frac{4m^2 h u^{-1}+\lambda {\cal T}}{g}.
\eeq
We find that only the magnetic part of the RR 3-form orthogonal to S$^2$ gives a Bianchi identity not implied by what has been derived thus far - this takes the form
\beq
g_{31}+C_0 H=\frac{1}{4m^2}\left(\frac{1}{2}\epsilon_{ijk}(\partial_{z_i}{\cal T}dy+\partial_{z_i}{\cal X} dx)\wedge dz_j\wedge dz_k+{\cal S} dz_{123}\right).
\eeq
The Bianchi identity of this flux components is implied when the RHS of this expression is closed, which requires that
\beq
\partial_y {\cal X}=\partial_x {\cal T},~~~~\partial_{z_i}^2{\cal T}=\partial_y {\cal S},~~~~\partial_{z_i}^2{\cal X}=\partial_x {\cal S},
\eeq
however only the last of these is not implied by the supersymmetry and Bianchi identity PDEs derived thus far.
\\
~~\\
In summary the class of solutions we derive in this section has an NS sector of the form
\begin{align}
ds^2&= \sqrt{\frac{g}{h}}u\bigg(ds^2(\text{AdS}_3)+\frac{1}{4m^2 \Xi}ds^2(\text{S}^2)\bigg)+\sqrt{\frac{h}{g}}\frac{dx^2}{u}+\frac{g^{\frac{3}{2}}}{4m^2\sqrt{h}}\left(dy+\frac{\lambda}{g}dx\right)^2+\frac{1}{4m^2}\sqrt{gh}(dz_i)^2,\nn\\[2mm]
e^{-\Phi}&=\frac{2m}{g}\sqrt{\frac{h \Xi}{u}},~~~2mH= d\left(\frac{g u u'}{
4m^2 h \Xi}-x\right)\wedge \text{vol}(\text{S}^2)-\frac{1}{2m}\bigg(\frac{1}{2}\epsilon_{ijk}(\partial_{z_i}g dy+\partial_{z_i}\lambda dx)\wedge dz_j\wedge dz_k- \partial_yh dz_{123}\bigg).\nn
\end{align}
It also has the following non trivial ten dimensional RR fluxes
\begin{align}
F_1&= dC_0,~~~~ C_0=w-\frac{\lambda}{g},~~~~F_5= (1+\star_{10})f_5\\[2mm]
F_3&=-C_0 H+\frac{1}{4m^2}\left(\frac{1}{2}\epsilon_{ijk}(\partial_{z_i}{\cal T}dy+\partial_{z_i}{\cal X} dx)\wedge dz_j\wedge dz_k+{\cal S} dz_{123}\right)\nn\\[2mm]
&-\frac{1}{2m}\left(d\left(\frac{{\cal T}u u'}{4m^2 h \Xi}+y\right)+w dx\right)\wedge\text{vol}(\text{S}^2),\nn\\[2mm]
f_5&=\frac{g^2}{2m^5 h\Xi}\bigg[\epsilon_{ijk}\left(\frac{uu'
}{2}\partial_{z_i}\left(\frac{\lambda}{g}\right)dy+\left(\frac{2m^2u'}{g}\partial_{z_i}\left(\frac{h}{g}\right)+\frac{uu'}{4}\partial_{z_i}\left(\frac{\lambda^2}{g^2}\right)\right)dx\right)dz_j\wedge dz_k\nn\\[2mm]
&+\bigg(\frac{4m^2 h^2}{g^2}-\frac{u^2 u'}{g}\partial_x(u^{-1}h)+\frac{u u' \lambda}{g}\partial_y h\bigg)dz_{123}\bigg]\wedge \text{vol}(\text{S}^2).\nn
\end{align}
Solutions in this class are in 1-to-1 correspondence with the solutions to the following PDEs: First supersymmetry demands that we impose
\beq
\partial_{y}\lambda=\partial_x g.
\eeq
Second, away form the loci of sources, the Bianchi identities impose 
\begin{align}
w'&= \partial_x\left(\frac{\lambda}{g}\right)-\frac{1}{2}\partial_y\left(\frac{\lambda^2}{g^2}\right)-\frac{4m^2 u}{g}\partial_y\left(\frac{h}{g^2 u^2}\right),\nn\\[2mm]
&\partial_{z_i}^2 g+ \partial_{y}^2h=0,~~~~~\partial_{z_i}^2 \lambda+ \partial_{x}\partial_{y}h=0,~~~~\partial_{z_i}^2{\cal X}=\partial_x {\cal S}.
\end{align}
Clearly the system of constraints that needs to be solved for this class is in general very complicated. This indicates two things: First that the class is likely quite broad, containing sub-classes with qualitatively different physics\footnote{A similarly complicated system derived for Mink$_4\times $S$^2$ in \cite{Macpherson:2016xwk} was shown to contain all (known) half BPS AdS$_{5,6,7}$ solutions modulo duality, as well as compact Minkowski vacua}; Second it suggests that there probably exists a better set of local coordinates on M$_5$ that simplifies these conditions some what. We have not made progress on the second point in general, but we  evidence the first point by deriving 2 simplified cases in the next section, assuming a diagonal metric ansatz.  

\subsubsection{2 interesting cases with diagonal metric}\label{eq:classIIrestriction}
The general $\sin\beta = 0$ class is rather complicated, so here we shall derive some simplified sub-classes. It is a generic feature of local classifications of supergravity solutions that those with the simplest PDES  governing them come with a diagonal metric, so this is what we shall pursue. The most obvious way to achieve this is to set $\lambda=0$, however it is possible to be more general than this: While it is true that the definition of \eqref{eq:caseIIviel} integrates $d(e^{A-\Phi}\cot\alpha K)\wedge dx$ without loss of generality, one could equally well take
\beq
e^{A-\Phi}\cos\alpha K = a(x,y)(b(y)dy+\hat{\lambda} dx),
\eeq
the difference is a diffeomorphism in $(x,y)$ that turns on $(a ,~b)$, setting $\hat\lambda=0$ here or $\tilde{\lambda}=0$ in \eqref{eq:caseIIviel} both give a diagonal metric, but the former is more general. We shall then take $(U,e_i)$ as defined in \eqref{eq:caseIIviel}, $\alpha,f,u$ as in  \eqref{eq:alphadef} and redefine the physical fields in terms of auxiliary functions as
\beq
e^{2A-\Phi}=2m\sqrt{\frac{au\tilde\Xi}{g}},~~~~e^{4A+3u}a=\frac{g u^2}{h},~~~~~\tilde{\Xi}= 1+\frac{g (u')^2}{4m^2 a h},~~~\hat\lambda =0,
\eeq
the last of these being the diagonal ansatz. Repeating the  same steps as the  previous section we find that supersymmetry requires  $\partial_x g=0$ and
\beq
4m^2H_3=-\frac{b}{2}\epsilon_{ijk}\partial_{z_i}g  dz_j\wedge dz_k\wedge dy+ \frac{1}{a b}\partial_y h dz_{123},~~~
f_1=b \partial_xa dy- \frac{4m^2 u}{bg}\partial_y\left(\frac{ag}{gu^2}\right)dx.
\eeq
In order for these flux components to satisfy their respective Bianchi identities we must have that
\beq\label{eq:branchsolll}
\partial_x\left(\frac{1}{a}\partial_y h  \right)=0,~~~~~~ \partial_{z_i}\bigg(\frac{1}{bg}\partial_y\left(\frac{ag}{gu^2}\right)\bigg)=0.
\eeq
These conditions do not have a unique general solution instead they represent a branching of possible classes of solution - ie  the solutions that follow from fixing $\partial_y\left(\frac{ag}{gu^2}\right)=0$ are distinct from those not obeying this constraint and so on. We have not found every branch that follows from \eqref{eq:branchsolll}, but shall provide 2 that result in  interesting physical systems. These follow from fixing the functions of our local ansatz as
\begin{align}
\text{case 1}:&~~ h=4m^2R T,~~~g =4m^2 T,~~~a=k,~~~b=1\\[1mm]
\text{case 2}:&~~ h=12m^2 T^2 S,~~~g=12\sqrt{3}m^2 T,~~~a=\sqrt{3}k^{\frac{5}{3}}\partial_y S,~~~b=\frac{1}{3} k^{\frac{4}{3}},\nn
\end{align}
where  the assumption comes with the choice of $(h,g,a)$, we are free to fix $b$ as we choose with diffeomorphisms. In all cases   $v=v(y)$ and $k=k(x)$, $R=R(x,z_i)$, $S=S(x,y)$  and $T=T(z_i)$. We will now present the cases that follow from these tunings\\[2mm]
~\\
\textbf{Case 1: A deformed D7-D5-NS5 brane intersection}\\
~\\
For case 1 $\partial_y$ is actually an isometry direction and  the local form of solutions reduces to
\begin{align}
ds^2&= \frac{u}{\sqrt{k R}}\bigg[ds^2(\text{AdS}_3)+\frac{1}{4m^2}\frac{1}{1+\frac{ (u')^2}{4m^2k R}}ds^2(\text{S}^2)\bigg]+\frac{\sqrt{k R }}{u} dx^2+T\bigg( \sqrt{\frac{k}{R}}dy^2 +\sqrt{\frac{R }{k}}(dz_i)^2\bigg)\nn\\[2mm]
e^{-\Phi}&=\frac{k \sqrt{R}}{\sqrt{T u}}\sqrt{1+\frac{ (u')^2}{4m^2k R}},~~~~2mH=d\left(\frac{u u'}{4m^2 k R\left(1+\frac{(u')^2}{4m^2 k R}\right)}-x\right)\wedge\text{vol}(\text{S}^2)- m \epsilon_{ijk}\partial_{z_i}T dz_j\wedge dz_k\wedge dy,\nn\\[2mm]
F_1&=k' dy,~~~~F_5=(1+\star_{10})f_5\\[2mm]
F_3&=\frac{1}{2m}\left(-k+\frac{uu' k'}{4m^2 k R+ (u')^2}\right)dy\wedge \text{vol}(\text{S}^2)+\frac{k}{2u}\epsilon_{ijk}\partial_{z_i}R dz_j\wedge dz_k\wedge dx-\partial_x R T dz_{123},\nn\\[2mm]
f_5&=\frac{1}{4m k R+(u')^2}\bigg[\frac{k u'}{4m}\epsilon_{ijk}\partial_{z_i}Rdz_j\wedge dz_k\wedge dx-\frac{1}{2m}T R \left(4m^2 k+ u' \partial_x(u R^{-1})\right)dz_{123}\bigg]\wedge \text{vol}(\text{S}^2)\nn.
\end{align}
Solutions in this case are defined entirely in terms of the Bianchi identities of the fluxes which impose the following system of PDEs away from the loci of sources
\beq\label{eq:sol1pdes}
k''=0,~~~~\partial_{z_i}^2 T=0,~~~~   k \partial_{z_i}^2 R+ u T \partial^2_xR=0.
\eeq
If we fix $u=k=1$ this is the system of PDEs of a flat space intersection of  D5 branes ending on NS5 branes that are smeared over $y$ \cite{Youm:1999ti}. The warp factors $T,R$  appear where one would expect if they are to be identified NS5 and D5 branes extended in AdS$_3\times$S$^2$ and $x$ or $y$ respectively. The effect of turning on $k$ is clearly to add D7 branes smeared on  $y$ to this system. When $u$ is non trivial, this cannot simply be interpreted as the warp factor of a brane, rather $u$ represents a deformation of this system, where the roles that $(R,T,k)$ play depend on the details of how \eqref{eq:sol1pdes} is solved. Finally we note T-dualising of $\partial_y$ maps us to a solution in massive IIA with an SU(2)$\times$ U(1) preserving squashed and fibred 3-sphere.\\
~~~~\\[2mm]
\textbf{Case 2: An Imamura-like case with D7-NS5-D5}\\
~\\
For case 2 not only is  $\partial_y$ not generically an isometry, it is manifestly never an isometry. The solutions in this case all take the local form
\begin{align}
ds^2&= \frac{u}{\sqrt{ S \partial_yS T}v^{\frac{5}{6}}}\bigg[ds^2(\text{AdS}_3)+\frac{1}{4m^2\tilde{\Xi}}ds^2(\text{S}^2)\bigg]+\frac{\sqrt{S\partial_yS T}}{v^{\frac{5}{6}} u }dx^2+\frac{T}{v^{\frac{5}{6}}}\bigg(\sqrt{\frac{\partial_y S}{ST}}\frac{dy^2}{v}+ 3\sqrt{\frac{S T}{\partial_y S}}(dz_i)^2\bigg),\nn\\[2mm]
e^{-\Phi}&= \frac{\sqrt{S}\partial_ySv^{\frac{5}{3}}}{\sqrt{3}\sqrt{u}}\sqrt{\tilde{\Xi}},~~~~\tilde{\Xi}=1+\frac{(u')^2}{4m^2 S\partial_yS T v^{\frac{5}{3}}}\\[2mm]
2m H&=d\left(\frac{u u'}{4m^2 S v^{\frac{5}{3}}T\tilde{\Xi}}-x\right)\wedge \text{vol}(\text{S}^2)-\frac{\sqrt{3}m}{v^{\frac{4}{3}}}\bigg(\epsilon_{ijk} \partial_{z_i}T dz_j\wedge dz_k- 8 v T^2 dz_{123}\bigg),\nn\\[2mm]
F_1&= \frac{1}{\sqrt{3}}d\left(v^{\frac{1}{3}}\partial_x S\right)+\frac{v'}{3 \sqrt{3}}\left(\frac{v^2 S\partial_yS}{u}dx- \frac{\partial_xS}{v^{\frac{2}{3}}}dy\right) ,~~~~~ F_5= (1+\star_{10})\tilde{f}_3\wedge \text{vol}(\text{S}^2),\nn\\[2mm]
F_3&=\frac{S\partial_y S v^{\frac{5}{3}}}{2 u}\epsilon_{ijk}\partial_{z_i}T dz_j\wedge dz_k\wedge dx-3 T^2\partial_x S dz_{123}-\frac{v^{\frac{1}{3}}}{2\sqrt{3}m}\partial_y S dy\wedge \text{vol}( \text{S}^2)\nn\\[2mm]
&+\frac{v^{\frac{1}{3}}}{8\sqrt{3}m^3   S\partial_y S T v^{\frac{5}{3}}\Xi}\bigg(u u' \partial_{x}\partial_y S dy+\frac{1}{3}u'( S\partial_y S v^{\frac{5}{3}} v'+3u \partial_x^2 S)\bigg)dx\wedge\text{vol}(\text{S}^2),\nn\\[2mm]
\tilde{f}_3&=\frac{1}{8m^3 S \partial_y S T v^{\frac{5}{3}}\tilde{\Xi}}\bigg[\frac{1}{2} S \partial_y S v^{\frac{5}{3}}u'\epsilon_{ijk}\partial_{z_i}T dz_j\wedge dz_k\wedge dx+3 T^2\left(4m^2 S\partial_y S T v^{\frac{5}{3}}+ S\partial_x(uu' S^{-1})\right)dz_{123}\bigg]\nn,
\end{align}
where $v$ is a non zero but otherwise arbitrary linear function of $y$. The Bianchi identities of the fluxes impose 
\beq
\partial_{z_i}^2 T=v' T^2,~~~~u\partial_x^2(\sqrt{v} S) + v^{\frac{13}{6}}\partial_y^2( ( \sqrt{v} S)^2)=0
\eeq
away from the loci of sources, there are no other conditions to be solved - we note that $v''=0$ is implied by the first of these, even in the presence of sources.  When $T=v=u=1$ the system of PDEs becomes very much like those of the D8-D6-NS5 system of \cite{Imamura:2001cr}, a $\frac{1}{4}$ BPS Mink$_6$ class in IIA  (see also \cite{Legramandi:2019ulq} for a version without an  SO(3) isometry). Of course here we are in IIB, so the interpretation must be a little different.  Examining the warp factors in the metric in this limit $\partial_y S$ appears to play the role of a localised D7 brane warp factor while $S$ that of D5 branes smeared on $z_i$. Turning on $T$ and defining $h_{D7}=\partial_yS$, $h_{\text{NS5}}= T$ and $h_{\text{D5}} = ST$ we see that these quantities appear in the metric where one would formally expect  the warp factors of a D7-D5-NS5 brane intersection to appear, however $S$ would need to be independent of $x$ for this interpretation. Additionally all these putative branes would have AdS$_3\times$ S$^2$ in their world volume and $z_i$ in their co dimensions, making this more closely resemble the  generalisation of \cite{Imamura:2001cr} presented in \cite{Legramandi:2019ulq}, T-dualised on a spacial direction in Mink$_6$ to get Mink$_5$ in IIB then compactified to AdS$_3\times$S$^2$. The effect of further turning on $(u,v)$ is to deform this system.  \\
~~\\
We have found that class II is really rather broad and that, even when rather draconian constraints are imposed on it, it is possible to extract interesting physical cases. We plan to explore the solutions contained here, in class II more broadly and in class I in forthcoming work \cite{NA}.

\section*{Acknowledgments}
NM would like to thank C. Couzens and A. Passias for collaboration on related projects. We thank Y. Lozano for interesting correspondences. NM  is supported by AEI-Spain (under project PID2020-114157GB-I00 and Unidad de Excelencia Mar\'\i a de Maetzu MDM-2016-0692), by Xunta de Galicia-Conseller\'\i a de Educaci\'on (Centro singular de investigaci\'on de Galicia accreditation 2019-2022, and project ED431C-2021/14), and by the European Union FEDER. AR is partially supported by the Spanish government grant PGC2018-096894-B-100. This research was supported in part by the National Science Foundation under Grant No. NSF PHY-1748958.
%%%%%%%%%%%%%%%%%%%%%%%%%%%%%%%%%%%%%%%%%%%%%%%%%%%%%%%%%%%%%%%%%%%%%%%%%%%%%%%%%%%%%%%%%%%%%%%%%%%%%%%%%%%%%%%%%%%%%%%%%%%%%%%%%%%%%%%%%%%%%%%%%%%%%%%%%%%%%%%%%%%%%%%%%%%%%%%%%%%%%%%%%%%%%%%%%%%%%%%%%%%%%%%%%%%%%%%%%%%%%%%%%%%%%%%%%%%%%%%%%%%%%%%%%%%%%%%%%%%%%%%%%%%%%%%%%%%%%%%%%%%%%%%%%%%%%%%%%%%%%%%%%%%%%%%%%%%%%%%%%%%%%%%%%%%%%%%%%%%%%%%%%%%%%%%%%%%%%%%%%%%%%%%%%%%%%%%%%%%%%%%%%%%%%%%%%%%%%%%%%%%%%%%%%%%%%%%%%%%%%%%%%%%%%%%%%%%%%%%%%%%%%%%%%%%%%%%%%%%%%%%%%%%%%%%%%%%%%%%%%%%%%%%%%%%%%%%%%%%%%%%%%%%%%%%%%%%%%%%%%%%%%%%%%%%%%%%%%%%%%%%%%%%%%%%%%%%%%%%%%%%%%%%%%%%%%%%%%%%%%%%%%%%
\appendix
\section{Gamma matrices on M$_7$ }\label{sec:gammamatrices}
To find the supersymmetric solutions given in the main text, we use the bispinor approach, thus, we start   
by taking a basis of 7d gamma matrices, $\gamma^{(7)}_A$,  that respect a 2+5 split (on S$^2\times$M$_5$) which decompose as,
\beq
\gamma^{(7)}_{i}=e^C \sigma_i\otimes \mathbb{I},~~~~\gamma^{(7)}_a=\sigma_3\otimes \gamma_a,~~~~\text{where}~~~~i\gamma_{1234567}^{(7)}=1. 
\eeq
Here, $\sigma_{i}$ (with $i=1,2$) are the gamma matrices on S$^2$ with chirality matrix $\sigma_3$, \textit{i.e.}  $\sigma_{1,2,3}$ are the Pauli matrices. In turn, $\gamma_a$ are the gamma matrices in 5d, namely $a=1,...,5$, such that the 5d intertwiner, $B_5$, satisfies
\beq
B_5B_5^*=-\mathbb{I},~~~~ B_5\gamma_aB_5^{-1}=\gamma_a^*.
\eeq
Finally, the intertwiner in 7d is defined as and satisfies
\beq
B^{(7)}= \sigma_2\otimes B_5~~~~\text{with}~~~~ B^{(7)}{B^{(7)}}^*=\mathbb{I},~~~~ B^{(7)}\gamma_A^{(7)}(B^{(7)})^{-1}=-{\gamma_A^{(7)}}^*
\eeq
and $A=1,...,7$.
\section{Geometric 5d conditions for supersymmetry}\label{sec:derivationof5dconstraints}
%\section{Bi-linears on M$_5$}
In this appendix we derive a set of geometric constraints for the five-dimensional manifold which are sufficient for supersymmetry. To be precise, we express these supersymmetric conditions in terms of five-dimensional bi-linears.
%In this appendix we give some details of the bi-linears
\subsection{Matrix bi-linears on M$_5$} \label{sec:appM2}
 We introduce four matrix bi-linears expressed in terms of the four 5d spinors, $\eta$, and their Majorana conjugates, namely 
 \beq
 (\psi_{\sigma\rho})^{\alpha\beta}=\eta^{\alpha}_{1\sigma}\otimes \eta^{\beta\dag}_{2\rho},~~~~\eta^{\alpha}= \left(\begin{array}{c}\eta\\ \eta^c\end{array}\right)^{\alpha}.
 \eeq
Note that,
 \begin{align}
 \eta^{\alpha}_{1}\otimes \eta^{\beta\dag}_{2}&=\left(\begin{array}{cc}\eta_1\otimes \eta_2^{\dag}& \eta_1\otimes \eta_2^{c\dag}\\(-\eta_1\otimes \eta_2^{c\dag})^*&(\eta_1\otimes \eta_2^{\dag})^*\end{array}\right)\\&= \text{Re}\eta_1\otimes \eta_2^{\dag}\delta^{ab}+i\text{Im}\eta_1\otimes \eta_2^{c\dag}(\sigma_1)^{ab}+i\text{Re}\eta_1\otimes\eta_2^{c\dag}(\sigma_2)^{\alpha\beta}+i \text{Im}\eta_1\otimes \eta_2^{\dag} (\sigma_3)^{\alpha\beta}\nn
 \end{align}
here the fundamental object is the 5 dimensional bi-spinor, which can be computed with following definition,
 \beq
 \label{bispinors}
\epsilon_1\otimes \epsilon^{\dag}_2= \frac{1}{2^{[d/2]}}\sum_{n=0}^d\frac{1}{n!}\epsilon_2^{\dag}\gamma_{a_1...a_n}\epsilon_1e^{a_1}\wedge...\wedge e^{a_n},
\eeq
in this case $\epsilon_1$ and $\epsilon_2$ are two d dimensional spinors, $\gamma_a$ a basis of the flat space gamma matrices in d dimensions and $e^{a}$ are the vielbein on the d dimensional space.\\ 
~\\
In the next sub-section we present details of the bi-spinors on S$^2$.
\subsection{Matrix bi-linears on S$^2$}\label{sec:appS2}
On S$^2$, there exist Killing spinors, $\xi$, which obey   
\beq
\nabla_{\mu}\xi=\frac{i}{2}\sigma_{\mu} \xi
\eeq 
where we used the Pauli matrices to represent the Clifford algebra on S$^2$,  $\mu=1,2$ is a flat index on the unit sphere and the 2d intertwiner defining Majorana conjugation is $\sigma_2$, such that $\xi^c=\sigma_2\xi^*$. As shown in \cite{Macpherson:2016xwk}, the  SU(2) doublets take the following form, 
\beq
\xi^{\alpha}= \left(\begin{array}{c}\xi\\ \xi^c\end{array}\right)^{\alpha},~~~ \sigma_3\xi^{\alpha}= \left(\begin{array}{c}\sigma_3\xi\\ \sigma_3\xi^c\end{array}\right)^{\alpha}
\eeq
where $\alpha$ and other Greek indices run over 1,2. We can form matrix bi-linears out of these doublets,
\begin{align}
\Xi^{\alpha\beta}=\xi^{\alpha}\otimes \xi^{\beta\dag},&~~~~\hat{\Xi}^{\alpha\beta}=\sigma_3\xi^{\alpha}\otimes \xi^{\beta\dag}, \\%&\sigma_3\xi^{\alpha}\otimes (\sigma_3\xi^{\beta})^{\dag}=(\xi^{\alpha}\otimes \xi^{\beta\dag})_+-(\xi^{\alpha}\otimes \xi^{\beta\dag})_-,~~~ \xi^{a}\otimes (\sigma_3\xi^{\beta})^{\dag}=(\sigma_3\xi^{\alpha}\otimes \xi^{\beta\dag})_+-(\sigma_3\xi^{\alpha}\otimes \xi^{\beta\dag})_-\nn,
\sigma_3\xi^{\alpha}\otimes (\sigma_3\xi^{\beta})^{\dag}=\Xi^{\alpha\beta}_+-\Xi^{\alpha\beta}_-,&~~~ \xi^{a}\otimes (\sigma_3\xi^{\beta})^{\dag}=\hat{\Xi}^{\alpha\beta}_+-\hat{\Xi}^{\alpha\beta}_-\nn,
\end{align}
here $\Xi^{\alpha\beta}_{\pm}$, $\hat{\Xi}^{\alpha\beta}_{\pm}$ are poly-forms containing only even/odd forms, which arise from the decomposition of $\Xi^{\alpha\beta}$ and $\hat{\Xi}^{\alpha\beta}$ via \eqref{bispinors}. 
In addition, $\Xi^{\alpha\beta}$ is linearly independent of $\hat{\Xi}^{\alpha\beta}$, component by component and at every form degree. Further one can show that, the only mixings of the S$^2$ bi-linears, which appear in the ${\cal N}=1$ supersymmetry constraints and are different to zero under d and wedge product are
\begin{align}
\label{dWedgeS2}
	&d\text{Re}\Xi^{\alpha\beta}_1=-2 \text{Im}\Xi^{\alpha\beta}_2,~~~d\text{Im}\Xi^{\alpha\beta}_1=2 \text{Re}\Xi^{\alpha\beta}_2,~~~d\text{Re}\hat{\Xi}^{\alpha\beta}_0=\text{Im}\hat{\Xi}^{\alpha\beta}_1,~~~d\text{Im}\hat{\Xi}^{\alpha\beta}_0=-\text{Re}\hat{\Xi}^{\alpha\beta}_1,\nn\\[2mm]
	&\Xi^{\alpha\beta}_0\text{vol}(\text{S}^2)=-\text{Im}\hat\Xi^{\alpha\beta}_2,~~~\text{Re}\hat{\Xi}^{\alpha\beta}_0\text{vol}(\text{S}^2)=-\text{Im}\Xi^{\alpha\beta}_2,~~~\text{Im}\hat{\Xi}^{\alpha\beta}_0\text{vol}(\text{S}^2)=\text{Re}\Xi^{\alpha\beta}_2.
\end{align}
and the only singlets are contained in the terms,
\beq
\label{singletsS2}
\Xi^{\alpha\beta}_0= \frac{1}{2}\delta^{\alpha\beta},~~~~\text{Im}\hat{\Xi}^{\alpha\beta}_2=-\frac{1}{2}\text{vol}(\text{S}^2)\delta^{\alpha\beta},
\eeq
which give rise to contributions to the RR fluxes.  Notice that $\text{Im}\Xi^{\alpha\beta}_0=\text{Re}\hat{\Xi}^{\alpha\beta}_2=0$. \\
~\\ In the next sub-section we will use the previous expressions to factor out the S$^2$ matrix bi-linears from the 7d bi-linears constraints leaving us with 5d conditions. 
%These facts are likely all that is needed as they should be enough to factor out the S$^2$ matrix bi-linears from the 7d bi-linears constraints leaving you with 5d conditions on the symmetric and anti symmetric parts of 5d matrix bi-linears.

%\subsection{5d matrix bi-linears}

\subsection{7d bi-linears as matrix bi-linear contractions}
In this section, we express the 7d bi-linears, as contractions of the S$^2$ and M$_5$ data.  
We take the representative ${\cal N}=1$ sub-sector of the ${\cal N}=4$ Majorana spinors as follows,
\beq
\chi_1= \frac{e^{\frac{A}{2}}}{\sqrt{2}}\bigg(\xi^{\alpha}\otimes\eta^{\alpha}_{11}+i \sigma_3\xi^{\alpha}\otimes\eta^{\alpha}_{12}\bigg),~~~\chi_2= \frac{e^{\frac{A}{2}}}{\sqrt{2}}\bigg(\xi^{\alpha}\otimes\eta^{\alpha}_{21}+i\sigma_3\xi^{\alpha}\otimes\eta^{\alpha}_{22}\bigg). 
\eeq 
Using the general feature that the bi-spinor can be decomposed as  
\begin{eqnarray}
\label{decomposition}
	\begin{split}
		\chi_1\otimes \chi_2^{\dag}=&\Psi_++i \Psi_-\\
		%=&\frac{e^{A}}{2}\bigg(\left[(\Psi_{11})^{ab}_++i(\Psi_{21})^{ab}_-\right]\wedge\Xi^{ab}+\left[(\Psi_{11})^{ab}_-+i(\Psi_{21})^{ab}_+\right]\wedge\tilde{\Xi}^{ab}\\
		%&+\left[(\Psi_{22})^{ab}_--i(\Psi_{12})^{ab}_+\right]\wedge[\tilde{\Xi}^{ab}_+-\tilde{\Xi}^{ab}_-]+\left[(\Psi_{22})^{ab}_+-i(\Psi_{12})^{ab}_-\right]\wedge[\Xi^{ab}_+-\Xi^{ab}_-]\bigg)
	\end{split}
\end{eqnarray}
and using the results established in the previous sections of this appendix, we can compute the following 7d bi-linears, 
%the what was established in the previous sections we can compute the 7d bi-linears,
%We find
%\begin{eqnarray}
%\begin{split}
%\Psi_+=\frac{e^A}{2}\left[({\Psi_{11}^{ab}}_{+}+\Psi^{ab}_{22+})\wedge\Xi_+^{ab}\red{+i}(\Psi^{ab}_{21-}+\Psi^{ab}_{12-})\wedge\Xi_1^{ab}+(\Psi^{ab}_{11-}-\Psi^{ab}_{22-})\wedge\tilde{\Xi}_1^{ab}\red{+i}(\Psi^{ab}_{21+}-\Psi^{ab}_{12+})\wedge\tilde{\Xi}_+^{ab}\right]\\
%i\Psi_-=\frac{e^A}{2}\left[({\Psi_{11}^{ab}}_{+}-\Psi^{ab}_{22+})\wedge\Xi%_1^{ab}\red{+i}(\Psi^{ab}_{21-}-\Psi^{ab}_{12-})\wedge\Xi_{+}^{ab}\red{+}(\Psi^{ab}_{11-}+\Psi^{ab}_{22-})\wedge\tilde{\Xi}_+^{ab}\red{+i}(\Psi^{ab}_{21+}+\Psi^{ab}_{12+})\wedge\tilde{\Xi}_1^{ab}\right]
%\end{split}
%\end{eqnarray}
%and expanding all the way out we find
\begin{subequations}
	\begin{align}
		\Psi_+=\frac{e^{A}}{2}\bigg(&\Xi^{\alpha\beta}_0\text{Re}(\psi^{(\alpha\beta)}_{11}+\psi^{(\alpha\beta)}_{22})_++e^{2C}\text{Im}\hat{\Xi}^{\alpha\beta}_2\wedge\text{Re}(\psi^{(\alpha\beta)}_{12}-\psi^{(\alpha\beta)}_{21})_++\text{Re}\hat\Xi^{\alpha\beta}_0\text{Im}(\psi^{(\alpha\beta)}_{12}-\psi^{(\alpha\beta)}_{21})_+\nn\\[2mm]
		+&\text{Im}\hat\Xi^{\alpha\beta}_0\text{Re}(\psi^{[\alpha\beta]}_{12}-\psi^{[\alpha\beta]}_{21})_++e^{C}\text{Re}\Xi^{\alpha\beta}_1\wedge\text{Im}(\psi^{(\alpha\beta)}_{12}+\psi^{(\alpha\beta)}_{21})_-+e^{C}\text{Im}\Xi^{\alpha\beta}_1\wedge\text{Re}(\psi^{[\alpha\beta]}_{12}+\psi^{[\alpha\beta]}_{21})_-\nn\\[2mm]
		-&e^{C}\text{Re}\hat{\Xi}^{\alpha\beta}_1\wedge\text{Re}(\psi^{[\alpha\beta]}_{11}-\psi^{[\alpha\beta]}_{22})_-+e^{C}\text{Im}\hat{\Xi}^{\alpha\beta}_1\wedge\text{Im}(\psi^{(\alpha\beta)}_{11}-\psi^{(\alpha\beta)}_{22})_-\nn\\[2mm]
		+&e^{2C}\text{Re}\Xi^{\alpha\beta}_2\wedge\text{Re}(\psi^{[\alpha\beta]}_{11}+\psi^{[\alpha\beta]}_{22})_+-e^{2C}\text{Im}\Xi^{\alpha\beta}_2\wedge\text{Im}(\psi^{(\alpha\beta)}_{11}+\psi^{(\alpha\beta)}_{22})_+\bigg),\label{eq:7dbilinears1}\\[2mm]
		\Psi_-=\frac{e^{A}}{2}\bigg(-&\Xi^{\alpha\beta}_0\text{Re}(\psi^{(\alpha\beta)}_{12}-\psi^{(\alpha\beta)}_{21})_-+e^{2C}\text{Im}\hat{\Xi}^{\alpha\beta}_2\wedge\text{Re}(\psi^{(\alpha\beta)}_{11}+\psi^{(\alpha\beta)}_{22})_-+\text{Re}\hat{\Xi}^{\alpha\beta}_0\text{Im}(\Psi_{11}^{(\alpha\beta)}+\Psi_{22}^{(\alpha\beta)})_-\nn\\[2mm]
		+&\text{Im}\hat{\Xi}^{\alpha\beta}_0\text{Re}(\Psi_{11}^{[\alpha\beta]}+\Psi_{22}^{[\alpha\beta]})_-+e^{C}\text{Re}\Xi^{\alpha\beta}_1\wedge\text{Im}(\psi^{(\alpha\beta)}_{11}-\psi^{(\alpha\beta)}_{22})_++e^{C}\text{Im}\Xi^{\alpha\beta}_1\wedge\text{Re}(\psi^{[\alpha\beta]}_{11}-\psi^{[\alpha\beta]}_{22})_+\nn\\[2mm]
		+&e^{C}\text{Re}\hat{\Xi}^{\alpha\beta}_1\wedge\text{Re}(\psi^{[\alpha\beta]}_{12}+\psi^{[\alpha\beta]}_{21})_+-e^{C}\text{Im}\hat{\Xi}^{\alpha\beta}_1\wedge\text{Im}(\psi^{(\alpha\beta)}_{12}+\psi^{(\alpha\beta)}_{21})_+\nn\\[2mm]
		-&e^{2C}\text{Re}\Xi^{\alpha\beta}_2\wedge\text{Re}(\psi^{[\alpha\beta]}_{12}-\psi^{[\alpha\beta]}_{21})_-+e^{2C}\text{Im}\Xi^{\alpha\beta}_2\wedge\text{Im}(\psi^{(\alpha\beta)}_{12}-\psi^{(\alpha\beta)}_{21})_-
		\bigg)\label{eq:7dbilinears2}
	\end{align}
\end{subequations}
in terms of the symmetric and antisymmetric parts of $\Psi_{ij}^{\alpha\beta}$. These expressions, along with \eqref{dWedgeS2}-\eqref{singletsS2}, will be useful in the next section, in the computations of the supersymmetric constraints. 

\subsection{From 7d to 5d supersymmetry constraints}\label{appendix:7dSUSY}
An ${\cal N}$=1 supersymmetric AdS$_3$ solution in type II supergravity, with purely magnetic NS flux, must obey the conditions 
 \cite{Passias:2020ubv}
\begin{subequations}
	\begin{align}
		&d_{H}(e^{A-\Phi}\Psi_{\mp})=0,\label{eq:bpsness1}\\[2mm]
		&d_{H}(e^{2A-\Phi}\Psi_{\pm})\mp 2m e^{A-\Phi}\Psi_{\mp}=\frac{e^{3A}}{8}\star_7\lambda(f),\label{eq:bpsness2}\\[2mm]
		&(\Psi_{\mp},f)_7= \mp \frac{m}{2} e^{-\Phi}\text{vol}(\text{M}_7)\label{eq:bpsness3},
	\end{align}
\end{subequations}
where $(\Psi_{\mp},f_{\pm})_7\equiv (\Psi_{\mp}\wedge \lambda(f_{\pm}))_7$ with $(X,Y)_7$ denoting the projection to the seven-form component. In \eqref{eq:bpsness1}-\eqref{eq:bpsness3}, an upper sign applies to type IIA and a lower one to type IIB. The twisted exterior derivative is defined as $d_H=d-H\wedge$, and in turn we are assuming a purely magnetic NS flux
\beq
H=e^{2C}H_1\wedge \text{vol}(\text{S}^2)+ H_3.
\eeq
Thus, one can write the RR flux poly-form term appearing in \eqref{eq:bpsness2} as following
\beq
\star_7\lambda f_{\pm}= -\star_5 \lambda g_{2\pm}+ e^{2C}\text{vol}(\text{S}^2)\wedge\star_5 \lambda g_{1\pm}= \delta^{\alpha\beta}(-\Xi^{\alpha\beta}\wedge \star_5 \lambda g_{2\pm}+i e^{2C} \hat{\Xi}^{\alpha\beta}\wedge \star_5\lambda g_{1\pm}).
\label{flux-gs}
\eeq

\subsubsection{IIA 5d conditions}
One can show that given the 7d bi-linears in \eqref{eq:7dbilinears1}-\eqref{eq:7dbilinears2} and making use of the expressions  \eqref{dWedgeS2} the supersymmetric constraints \eqref{eq:bpsness1}-\eqref{eq:bpsness2}, independent of the RR forms,  are equivalent to the following conditions in 5d, 
\begin{subequations}
	\begin{align}
		&d_{H_3}\big(e^{2A-\Phi}\text{Re}(\psi^{(\alpha\beta)}_{12}-\psi^{(\alpha\beta)}_{21})_-\big)=0,\label{eq:IIfirstbispinorconds1}\\[2mm]
		&d_{H_3}\big(e^{2A+C-\Phi}\text{Re}(\psi^{[\alpha\beta]}_{12}+\psi^{[\alpha\beta]}_{21})_+\big)+e^{2A-\Phi}\text{Re}(\psi^{[\alpha\beta]}_{11}+\psi^{[\alpha\beta]}_{22})_-=0,\label{eq:IIfirstbispinorconds2}\\[2mm]
		&d_{H_3}\big(e^{2A+C-\Phi}\text{Im}(\psi^{(\alpha\beta)}_{12}+\psi^{(\alpha\beta)}_{21})_+\big)+e^{2A-\Phi}\text{Im}(\psi^{(\alpha\beta)}_{11}+\psi^{(\alpha\beta)}_{22})_-=0,\label{eq:IIfirstbispinorconds3}\\[2mm]
		&d_{H_3}\big(e^{2A+2C-\Phi}\text{Re}(\psi^{(\alpha\beta)}_{11}+\psi^{(\alpha\beta)}_{22})_-\big)- e^{2A+2C-\Phi} H_1\wedge \text{Re}(\psi^{(\alpha\beta)}_{12}-\psi^{(\alpha\beta)}_{21})_-=0,\label{eq:IIfirstbispinorconds4}\\[2mm]
		&d_{H_3}\big(e^{2A+2C-\Phi}\text{Im}(\psi^{(\alpha\beta)}_{12}-\psi^{(\alpha\beta)}_{21})_-\big)+ e^{2A+2C-\Phi} H_1\wedge \text{Im}(\psi^{(\alpha\beta)}_{11}+\psi^{(\alpha\beta)}_{22})_--2 e^{2A+C-\Phi}\text{Im}(\psi^{(\alpha\beta)}_{11}-\psi^{(\alpha\beta)}_{22})_+=0,\label{eq:IIfirstbispinorconds5}\\[2mm]
		&d_{H_3}\big(e^{2A+2C-\Phi}\text{Re}(\psi^{[\alpha\beta]}_{12}-\psi^{[\alpha\beta]}_{21})_-\big)+ e^{2A+2C-\Phi} H_1\wedge \text{Re}(\psi^{[\alpha\beta]}_{11}+\psi^{[\alpha\beta]}_{22})_-- 2 e^{2A+C-\Phi}\text{Re}(\psi^{[\alpha\beta]}_{11}-\psi^{[\alpha\beta]}_{22})_+=0,\label{eq:IIfirstbispinorconds6}\\
		%&\color{blue}{d_{H_3}\big(e^{3A-\Phi}\text{Im}(\Psi^{(\alpha\beta)}_{12}-\Psi^{(\alpha\beta)}_{21})_+\big)-2me^{2A-\Phi}\text{Im}(\Psi^{(\alpha\beta)}_{11}+\Psi^{(\alpha\beta)}_{22})_-=0,}\label{eq:IIsecondbispinorconds1}\\[2mm]
		%&\color{blue}{d_{H_3}\big(e^{3A-\Phi}\text{Re}(\Psi^{[\alpha\beta]}_{12}-\Psi^{[\alpha\beta]}_{21})_+\big)-2me^{2A-\Phi}\text{Re}(\Psi^{[\alpha\beta]}_{11}+\Psi^{[\alpha\beta]}_{22})_-=0,}\label{eq:IIsecondbispinorconds2}\\[2mm]
		%&\color{blue}{d_{H_3}\big(e^{3A+C-\Phi}\text{Im}(\Psi^{(\alpha\beta)}_{12}+\Psi^{(\alpha\beta)}_{21})_-\big)+2me^{2A+C-\Phi}\text{Im}(\Psi^{(\alpha\beta)}_{11}-\Psi^{(\alpha\beta)}_{22})_+=0,}\label{eq:IIsecondbispinorconds3}\\[2mm]
		%&\color{blue}{d_{H_3}\big(e^{3A+C-\Phi}\text{Re}(\Psi^{[\alpha\beta]}_{12}+\Psi^{[\alpha\beta]}_{21})_-\big)+2me^{2A+C-\Phi}\text{Re}(\Psi^{[\alpha\beta]}_{11}-\Psi^{[\alpha\beta]}_{22})_+=0,}\label{eq:IIsecondbispinorconds4}\\[2mm]
		&d_{H_3}\big(e^{3A+C-\Phi}\text{Re}(\psi^{[\alpha\beta]}_{11}-\psi^{[\alpha\beta]}_{22})_-\big)-e^{3A-\Phi}\text{Re}(\psi^{[\alpha\beta]}_{12}-\psi^{[\alpha\beta]}_{21})_+-2me^{2A+C-\Phi}\text{Re}(\psi^{[\alpha\beta]}_{12}+\psi^{[\alpha\beta]}_{21})_+=0,\label{eq:IIsecondbispinorconds5}\\[2mm]
		&d_{H_3}\big(e^{3A+C-\Phi}\text{Im}(\psi^{(\alpha\beta)}_{11}-\psi^{(\alpha\beta)}_{22})_-\big)-e^{3A-\Phi}\text{Im}(\psi^{(\alpha\beta)}_{12}-\psi^{(\alpha\beta)}_{21})_+-2me^{2A+C-\Phi}\text{Im}(\psi^{(\alpha\beta)}_{12}+\psi^{(\alpha\beta)}_{21})_+=0,\label{eq:IIsecondbispinorconds6}\\[2mm]
		&d_{H_3}\big(e^{3A+2C-\Phi}\text{Re}(\psi^{[\alpha\beta]}_{11}+\psi^{[\alpha\beta]}_{22})_+\big)+2e^{3A+C-\Phi}\text{Re}(\psi^{[\alpha\beta]}_{12}+\psi^{[\alpha\beta]}_{21})_-+2me^{2A+2C-\Phi}\text{Re}(\psi^{[\alpha\beta]}_{12}-\psi^{[\alpha\beta]}_{21})_-\nonumber\\[2mm]
		&~~~~~~~~~~~~~~~~~~~~-e^{3A+2C-\Phi}H_1\wedge\text{Re}(\psi^{[\alpha\beta]}_{12}-\psi^{[\alpha\beta]}_{21})_+=0,\label{eq:IIsecondbispinorconds7}\\[2mm]
		&d_{H_3}\big(e^{3A+2C-\Phi}\text{Im}(\psi^{(\alpha\beta)}_{11}+\psi^{(\alpha\beta)}_{22})_+\big)+2e^{3A+C-\Phi}\text{Im}(\psi^{(\alpha\beta)}_{12}+\psi^{(\alpha\beta)}_{21})_-+2me^{2A+2C-\Phi}\text{Im}(\psi^{(\alpha\beta)}_{12}-\psi^{(\alpha\beta)}_{21})_-\nonumber\\[2mm]
		&~~~~~~~~~~~~~~~~~~~~-e^{3A+2C-\Phi}H_1\wedge\text{Im}(\psi^{(\alpha\beta)}_{12}-\psi^{(\alpha\beta)}_{21})_+=0,\label{eq:IIsecondbispinorconds8}
	\end{align} 
	\label{eq:firstsecond}
\end{subequations}
and some more that are equivalent to,
\begin{subequations}
\label{otherEqs}
	\begin{align}
		&\text{Re}(\psi^{[\alpha\beta]}_{12}+\psi^{[\alpha\beta]}_{21})_+\wedge dH_3=\text{Im}(\psi^{(\alpha\beta)}_{12}+\psi^{(\alpha\beta)}_{21})_+\wedge dH_3=0,\\[2mm]
		&e^{2C}\text{Im}(\psi^{(\alpha\beta)}_{12}-\psi^{(\alpha\beta)}_{21})_-\wedge dH_3-d(e^{2C}H_1)\wedge \text{Im}(\psi^{(\alpha\beta)}_{11}+\psi^{(\alpha\beta)}_{22})_-=0,\\[2mm]
		&e^{2C}\text{Re}(\psi^{[\alpha\beta]}_{12}-\psi^{[\alpha\beta]}_{21})_-\wedge dH_3-d(e^{2C}H_1)\wedge \text{Re}(\psi^{[\alpha\beta]}_{11}+\psi^{[\alpha\beta]}_{22})_-=0\\
		&\text{Re}(\psi^{[\alpha\beta]}_{11}-\psi^{[\alpha\beta]}_{22})_-\wedge dH_3=\text{Im}(\psi^{(\alpha\beta)}_{11}-\psi^{(\alpha\beta)}_{22})_-\wedge dH_3=0,\\[2mm]
		&e^{2C}\text{Re}(\psi^{[\alpha\beta]}_{11}+\psi^{[\alpha\beta]}_{22})_+\wedge dH_3+d(e^{2C}H_1)\wedge \text{Re}(\psi^{[\alpha\beta]}_{12}-\psi^{[\alpha\beta]}_{21})_+=0,\\[2mm]
		&e^{2C}\text{Im}(\psi^{(\alpha\beta)}_{11}+\psi^{(\alpha\beta)}_{22})_+\wedge dH_3+d(e^{2C}H_1)\wedge \text{Im}(\psi^{(\alpha\beta)}_{12}-\psi^{(\alpha\beta)}_{21})_+=0.
	\end{align}
\end{subequations}
These previous expressions are implied when the sourceless Bianchi identity of the NS flux is obeyed and more broadly constrain exactly what source terms $dH_3$ and d$e^{2C}H_1$ we can have.\\
~\\
Using \eqref{flux-gs}, the RR field-strengths are derived from \eqref{eq:7dbilinears2},
\begin{subequations}
	\begin{align}
		e^{3A}\delta^{\alpha\beta}\star_5\lambda g_2&=-4d_{H_3}\big(e^{3A-\Phi}\text{Re}(\psi^{(\alpha\beta)}_{11}+\psi^{(\alpha\beta)}_{22})_+\big)-8me^{2A-\Phi}\text{Re}(\psi^{(\alpha\beta)}_{12}-\psi^{(\alpha\beta)}_{21})_-,\\[2mm]
		e^{3A+2C}\delta^{\alpha\beta}\star_5\lambda g_1&=-4d_{H_3}\big(e^{3A+2C-\Phi}\text{Re}(\psi^{(\alpha\beta)}_{12}-\psi^{(\alpha\beta)}_{21})_+\big)+8me^{2A+2C-\Phi}\text{Re}(\psi^{(\alpha\beta)}_{11}+\psi^{(\alpha\beta)}_{22})_-\nonumber\\[2mm]&~~~~~~~~~~~~~~~~~~~~~~~~~~~~~~~~~~~~~~~~~~~~
		-4e^{3A+2C-\Phi}H_1\wedge\text{Re}(\psi^{(\alpha\beta)}_{11}+\psi^{(\alpha\beta)}_{22})_+.
	\end{align}
	\label{fluxeqs}
\end{subequations}
Finally, considering $\lambda(f_+)=-\lambda g_{1+}-e^{2C}\text{vol}(\text{S}^2)\wedge\lambda g_{2+}$ in the pairing equation \eqref{eq:bpsness3} yields the additional restrictions,
\begin{subequations}
	\begin{align}
		&\delta^{\alpha\beta}\big(\text{Re}\big(\psi_{11}^{(\alpha\beta)}+\psi_{22}^{(\alpha\beta)}\big)_{-}\wedge\lambda g_{1+}+\text{Re}\big(\psi_{12}^{(\alpha\beta)}-\psi_{21}^{(\alpha\beta)}\big)_{-}\wedge\lambda g_{2+}\big)=-2\mu e^{-(\Phi+A)}\text{vol}(\text{M}_5),\label{third1}\\[2mm]
		&\text{Im}\big(\psi_{11}^{(\alpha\beta)}+\psi_{22}^{(\alpha\beta)}\big)_{-}\wedge\lambda g_{2+}=\text{Im}\big(\psi_{12}^{(\alpha\beta)}-\psi_{21}^{(\alpha\beta)}\big)_{-}\wedge\lambda g_{1+},\label{third2}\\[2mm]
		&\text{Re}\big(\psi_{11}^{[\alpha\beta]}+\psi_{22}^{[\alpha\beta]}\big)_{-}\wedge\lambda g_{2+}=\text{Re}\big(\psi_{12}^{[\alpha\beta]}-\psi_{21}^{[\alpha\beta]}\big)_{-}\wedge\lambda g_{1+}\label{third3}.
	\end{align}
	\label{thirdgen} 
\end{subequations}
The equations \eqref{eq:firstsecond}, \eqref{otherEqs},\eqref{fluxeqs}, \eqref{thirdgen} are sufficient constraints for the preservation of supersymmetry in type IIA.\\
~\\
Some particularly important conditions are the matrix 0-form constraints coming from \eqref{eq:firstsecond}, as these are purely algebraic namely
\beq
\begin{split}
&\eta_{21}^{c\dag}\eta_{11}=\eta_{22}^{c\dag}\eta_{12},\qquad \text{Im}(\eta_{21}^{\dag}\eta_{11})=\text{Im}(\eta_{22}^{\dag}\eta_{12}),\\
&(1+2me^{C-A})\eta_{22}^{c\dag}\eta_{11}=(1-2me^{C-A})\eta_{21}^{c\dag}\eta_{12},\\
&(1+2me^{C-A})\text{Im}(\eta_{22}^{\dag}\eta_{11})=(1-2me^{C-A})\text{Im}(\eta_{21}^{\dag}\eta_{12}).
\end{split}
\eeq
We will use these to constrain the spinors in the main text. 
\subsubsection{IIB 5d conditions}
In IIB, the supersymmetric constraints \eqref{eq:bpsness1}-\eqref{eq:bpsness2}, independent of the RR forms, are equivalent to the following conditions in 5d, 
\begin{subequations}
\begin{align}
&d_{H_3}\big(e^{2A-\Phi}\text{Re}(\Psi^{(\alpha\beta)}_{11}+\Psi^{(\alpha\beta)}_{22})_+\big)=0,\label{5deqIIB1}\\[2mm]
%&\color{blue}{d_{H_3}\big(e^{2A-\Phi}\text{Im}(\Psi^{(\alpha\beta)}_{12}-\Psi^{(\alpha\beta)}_{21})_+\big)=0},\\[2mm]
%&\color{blue}{d_{H_3}\big(e^{2A-\Phi}\text{Re}(\Psi^{[\alpha\beta]}_{12}-\Psi^{[\alpha\beta]}_{21})_+\big)=0},\\[2mm]
%&\color{blue}{d_{H_3}\big(e^{2A+C-\Phi}\text{Im}(\Psi^{(\alpha\beta)}_{12}+\Psi^{(\alpha\beta)}_{21})_-\big)=0},\\[2mm]
%&\color{blue}{d_{H_3}\big(e^{2A+C-\Phi}\text{Re}(\Psi^{[\alpha\beta]}_{12}+\Psi^{[\alpha\beta]}_{21})_-\big)=0},\\[2mm]
&d_{H_3}\big(e^{2A+C-\Phi}\text{Re}(\Psi^{[\alpha\beta]}_{11}-\Psi^{[\alpha\beta]}_{22})_-\big)-e^{2A-\Phi}\text{Re}(\Psi^{[\alpha\beta]}_{12}-\Psi^{[\alpha\beta]}_{21})_+=0,\label{5deqIIB2}\\[2mm]
&d_{H_3}\big(e^{2A+C-\Phi}\text{Im}(\Psi^{(\alpha\beta)}_{11}-\Psi^{(\alpha\beta)}_{22})_-\big)-e^{2A-\Phi}\text{Im}(\Psi^{(\alpha\beta)}_{12}-\Psi^{(\alpha\beta)}_{21})_+=0,\label{5deqIIB3}\\[2mm]
&d_{H_3}\big(e^{2A+2C-\Phi}\text{Re}(\Psi^{[\alpha\beta]}_{11}+\Psi^{[\alpha\beta]}_{22})_+\big)-e^{2A+2C-\Phi} H_1\wedge \text{Re}(\Psi^{[\alpha\beta]}_{12}-\Psi^{[\alpha\beta]}_{21})_++2e^{2A+C-\Phi}\text{Re}(\Psi^{[\alpha\beta]}_{12}+\Psi^{[\alpha\beta]}_{21})_-=0,\label{5deqIIB4}\\[2mm]
&d_{H_3}\big(e^{2A+2C-\Phi}\text{Im}(\Psi^{(\alpha\beta)}_{11}+\Psi^{(\alpha\beta)}_{22})_+\big)- e^{2A+2C-\Phi} H_1\wedge \text{Im}(\Psi^{(\alpha\beta)}_{12}-\Psi^{(\alpha\beta)}_{21})_++2 e^{2A+C-\Phi}\text{Im}(\Psi^{(\alpha\beta)}_{12}+\Psi^{(\alpha\beta)}_{21})_-=0,\label{5deqIIB5}\\[2mm]
&d_{H_3}\big(e^{2A+2C-\Phi}\text{Re}(\Psi^{(\alpha\beta)}_{12}-\Psi^{(\alpha\beta)}_{21})_+\big)+ e^{2A+2C-\Phi} H_1\wedge \text{Re}(\Psi^{(\alpha\beta)}_{11}+\Psi^{(\alpha\beta)}_{22})_+=0,\label{5deqIIB6}\\[2mm]
%&\color{blue}{d_{H_3}\big(e^{3A-\Phi}\text{Im}(\Psi^{(\alpha\beta)}_{11}+\Psi^{(\alpha\beta)}_{22})_-\big)+2me^{2A-\Phi}\text{Im}(\Psi^{(\alpha\beta)}_{12}-\Psi^{(\alpha\beta)}_{21})_+=0,}\\[2mm]
%&\color{blue}{d_{H_3}\big(e^{3A-\Phi}\text{Re}(\Psi^{[\alpha\beta]}_{11}+\Psi^{[\alpha\beta]}_{22})_-\big)+2me^{2A-\Phi}\text{Re}(\Psi^{[\alpha\beta]}_{12}-\Psi^{[\alpha\beta]}_{21})_+=0,}\\[2mm]
%&\color{blue}{d_{H_3}\big(e^{3A+C-\Phi}\text{Im}(\Psi^{(\alpha\beta)}_{11}-\Psi^{(\alpha\beta)}_{22})_+\big)-2me^{2A+C-\Phi}\text{Im}(\Psi^{(\alpha\beta)}_{12}+\Psi^{(\alpha\beta)}_{21})_-=0,}\\[2mm]
%&\color{blue}{d_{H_3}\big(e^{3A+C-\Phi}\text{Re}(\Psi^{[\alpha\beta]}_{11}-\Psi^{[\alpha\beta]}_{22})_+\big)-2me^{2A+C-\Phi}\text{Re}(\Psi^{[\alpha\beta]}_{12}+\Psi^{[\alpha\beta]}_{21})_-=0,}\\[2mm]
&d_{H_3}\big(e^{3A+C-\Phi}\text{Re}(\Psi^{[\alpha\beta]}_{12}+\Psi^{[\alpha\beta]}_{21})_+\big)+e^{3A-\Phi}\text{Re}(\Psi^{[\alpha\beta]}_{11}+\Psi^{[\alpha\beta]}_{22})_-+2me^{2A+C-\Phi}\text{Re}(\Psi^{[\alpha\beta]}_{11}-\Psi^{[\alpha\beta]}_{22})_-=0,\label{5deqIIB7}\\[2mm]
&d_{H_3}\big(e^{3A+C-\Phi}\text{Im}(\Psi^{(\alpha\beta)}_{12}+\Psi^{(\alpha\beta)}_{21})_+\big)+e^{3A-\Phi}\text{Im}(\Psi^{(\alpha\beta)}_{11}+\Psi^{(\alpha\beta)}_{22})_-+2me^{2A+C-\Phi}\text{Im}(\Psi^{(\alpha\beta)}_{11}-\Psi^{(\alpha\beta)}_{22})_-=0,\label{5deqIIB8}\\[2mm]
&d_{H_3}\big(e^{3A+2C-\Phi}\text{Re}(\Psi^{[\alpha\beta]}_{12}-\Psi^{[\alpha\beta]}_{21})_-\big)-2e^{3A+C-\Phi}\text{Re}(\Psi^{[\alpha\beta]}_{11}-\Psi^{[\alpha\beta]}_{22})_+-2me^{2A+2C-\Phi}\text{Re}(\Psi^{[\alpha\beta]}_{11}+\Psi^{[\alpha\beta]}_{22})_+\nonumber\\[2mm]
&~~~~~~~~~~~~~~~+e^{3A+2C-\Phi}H_1\wedge\text{Re}(\Psi^{[\alpha\beta]}_{11}+\Psi^{[\alpha\beta]}_{22})_-=0,\label{5deqIIB9}\\[2mm]
&d_{H_3}\big(e^{3A+2C-\Phi}\text{Im}(\Psi^{(\alpha\beta)}_{12}-\Psi^{(\alpha\beta)}_{21})_-\big)-2e^{3A+C-\Phi}\text{Im}(\Psi^{(\alpha\beta)}_{11}-\Psi^{(\alpha\beta)}_{22})_+-2me^{2A+2C-\Phi}\text{Im}(\Psi^{(\alpha\beta)}_{11}+\Psi^{(\alpha\beta)}_{22})_+\nonumber\\[2mm]
&~~~~~~~~~~~~~~~+e^{3A+2C-\Phi}H_1\wedge\text{Im}(\Psi^{(\alpha\beta)}_{11}+\Psi^{(\alpha\beta)}_{22})_-=0\label{5deqIIB10}
\end{align}
\label{5deqIIB}
\end{subequations}
and some more expressions, which are implied when the sourceless Bianchi identity of the NS flux is obeyed,
 that are equivalent to,
\begin{subequations}
\begin{align}
&\text{Re}(\Psi^{[\alpha\beta]}_{11}-\Psi^{[\alpha\beta]}_{22})_-\wedge dH_3=\text{Im}(\Psi^{(\alpha\beta)}_{11}-\Psi^{(\alpha\beta)}_{22})_-\wedge dH_3=0,\\[2mm]
&e^{2C}\text{Im}(\Psi^{(\alpha\beta)}_{11}+\Psi^{(\alpha\beta)}_{22})_+\wedge dH_3+d(e^{2C}H_1)\wedge \text{Im}(\Psi^{(\alpha\beta)}_{12}-\Psi^{(\alpha\beta)}_{21})_+=0,\label{5deqIIB11}\\[2mm]
&e^{2C}\text{Re}(\Psi^{[\alpha\beta]}_{11}+\Psi^{[\alpha\beta]}_{22})_+\wedge dH_3+d(e^{2C}H_1)\wedge \text{Re}(\Psi^{[\alpha\beta]}_{12}-\Psi^{[\alpha\beta]}_{21})_+=0,\label{5deqIIB12}\\[2mm]
&\text{Re}(\Psi^{[\alpha\beta]}_{12}+\Psi^{[\alpha\beta]}_{21})_+\wedge dH_3=\text{Im}(\Psi^{(\alpha\beta)}_{12}+\Psi^{(\alpha\beta)}_{21})_+\wedge dH_3=0,\\[2mm]
&e^{2C}\text{Re}(\Psi^{[\alpha\beta]}_{12}-\Psi^{[\alpha\beta]}_{21})_-\wedge dH_3-d(e^{2C}H_1)\wedge \text{Re}(\Psi^{[\alpha\beta]}_{11}+\Psi^{[\alpha\beta]}_{22})_-=0,\label{5deqIIB13}\\[2mm]
&e^{2C}\text{Im}(\Psi^{(\alpha\beta)}_{12}-\Psi^{(\alpha\beta)}_{21})_-\wedge dH_3-d(e^{2C}H_1)\wedge \text{Im}(\Psi^{(\alpha\beta)}_{11}+\Psi^{(\alpha\beta)}_{22})_-=0.\label{5deqIIB14}
\end{align}
\end{subequations}
In turn, the flux equations derive from \eqref{eq:7dbilinears1} and \eqref{flux-gs} are,
\begin{subequations}
\begin{align}
e^{3A}\delta^{\alpha\beta}\star_5\lambda g_2&=4d_{H_3}\big(e^{3A-\Phi}\text{Re}(\Psi^{(\alpha\beta)}_{12}-\Psi^{(\alpha\beta)}_{21})_-\big)-8me^{2A-\Phi}\text{Re}(\Psi^{(\alpha\beta)}_{11}+\Psi^{(\alpha\beta)}_{22})_+,\label{5deqIIB15}\\[2mm]
e^{3A+2C}\delta^{\alpha\beta}\star_5\lambda g_1&=-4d_{H_3}\big(e^{3A+2C-\Phi}\text{Re}(\Psi^{(\alpha\beta)}_{11}+\Psi^{(\alpha\beta)}_{22})_-\big)-8me^{2A+2C-\Phi}\text{Re}(\Psi^{(\alpha\beta)}_{12}-\Psi^{(\alpha\beta)}_{21})_+\nonumber\\[2mm]
&~~~~~~~~~~~~~~~~~~~~~~~~~~~~~~~~~~~~~~~~~~~~+4e^{3A+2C-\Phi}H_1\wedge\text{Re}(\Psi^{(\alpha\beta)}_{12}-\Psi^{(\alpha\beta)}_{21})_-\label{5deqIIB16}
\end{align}
\end{subequations}
and considering $\lambda(f_-)=-\lambda g_{1-}-e^{2C}\text{vol}(\text{S}^2)\wedge\lambda g_{2-}$ in the pairing equation \eqref{eq:bpsness3} we obtain the following restrictions,
\begin{subequations}
	\begin{align}
		&\delta^{\alpha\beta}\big(\text{Re}\big(\psi_{11}^{(\alpha\beta)}+\psi_{22}^{(\alpha\beta)}\big)_{+}\wedge\lambda g_{2-}-\text{Re}\big(\psi_{12}^{(\alpha\beta)}-\psi_{21}^{(\alpha\beta)}\big)_{+}\wedge\lambda g_{1-}\big)=-2\mu e^{-(\Phi+A)}\text{vol}(\text{M}_5),\label{third1IIB}\\[2mm]
		&\text{Im}\big(\psi_{12}^{(\alpha\beta)}-\psi_{21}^{(\alpha\beta)}\big)_{+}\wedge\lambda g_{2-}=-\text{Im}\big(\psi_{11}^{(\alpha\beta)}+\psi_{22}^{(\alpha\beta)}\big)_{+}\wedge\lambda g_{1-},\label{third2IIB}\\[2mm]
		&\text{Re}\big(\psi_{12}^{[\alpha\beta]}-\psi_{21}^{[\alpha\beta]}\big)_{+}\wedge\lambda g_{2-}=-\text{Re}\big(\psi_{11}^{[\alpha\beta]}+\psi_{22}^{[\alpha\beta]}\big)_{+}\wedge\lambda g_{1-}\label{third3IIB}.
	\end{align}
	\label{thirdgenIIB} 
\end{subequations}
The matrix 0-form constraints coming from \eqref{5deqIIB} are,
\beq
\begin{split}
&\eta_{22}^{c\dag}\eta_{11}=\eta_{21}^{c\dag}\eta_{12},\qquad \text{Im}(\eta_{22}^{\dag}\eta_{11})=\text{Im}(\eta_{21}^{\dag}\eta_{12}),\\
&(1+me^{C-A})\eta_{21}^{\dag}\eta_{11}=(1-me^{C-A})\eta_{22}^{\dag}\eta_{12},\\
&(1+me^{C-A})\text{Im}(\eta_{21}^{c\dag}\eta_{11})=(1-me^{C-A})\text{Im}(\eta_{22}^{c\dag}\eta_{12}).
\end{split}
\eeq
which will be used in the main text in order to find restrictions in the spinors.

\end{document}